\documentclass[amsmath,amssymb,11pt]{article}
\usepackage{jheppub2}
\pdfoutput=1

\usepackage{amsmath,amssymb,amsthm,mathrsfs,bbm,bm}
\usepackage{latexsym,amscd,amsbsy,amsfonts,dsfont}
\usepackage{multirow}
\usepackage{graphicx}
\usepackage{xcolor}
\usepackage{caption}

\newcommand{\beq}{\begin{equation}}
\newcommand{\eeq}{\end{equation}}
\newcommand{\beqa}{\begin{eqnarray}}
\newcommand{\eeqa}{\end{eqnarray}}
\newcommand{\bea}{\begin{eqnarray}}
\newcommand{\eea}{\end{eqnarray}}

\newcommand*{\affmark}[1][*]{\textsuperscript{#1}} % For the affiliations

\numberwithin{equation}{section}  % make eq labels (sec.num)
\allowdisplaybreaks

\title{Quantum Transparency of Near-extremal Black Holes}

\author{Roberto Emparan\affmark[1,2],}
\emailAdd{emparan@ub.edu}
\author{Stefano Trezzi\affmark[2]}
\emailAdd{strezzi@icc.ub.edu}

\affiliation{
\affmark[1]Institució Catalana de Recerca i Estudis Avançats (ICREA),
 Passeig Lluis Companys, 23, 08010 Barcelona, Spain\\
\affmark[2]Departament de Física Quàntica i Astrofísica and
  Institut de Ciències del Cosmos, Martì i Franquès 1,
 Universitat de Barcelona, 08028 Barcelona, Spain\\
}

\abstract{We investigate the scattering of electromagnetic and gravitational waves off a Reissner-Nordstr\"om black hole in the low-temperature regime where the near-horizon throat experiences large quantum fluctuations. We find that the black hole is transparent to electromagnetic and gravitational radiation of fixed helicity below a certain frequency threshold. This phenomenon arises because the angular momentum of the black hole is quantized, creating an energy gap between the spinless black hole state and the first excited spinning states. Radiation with angular momentum---such as photons, gravitons, and partial waves of a massless scalar field, which we also study---must supply enough energy to bridge this gap to be absorbed. Below this threshold, no absorption can occur, rendering the black hole transparent. For frequencies above the gap, the scarcity of black hole states continues to suppress the absorption cross-section relative to semiclassical predictions, making the black hole translucent rather than completely transparent. Notably, electromagnetic absorption is significantly stronger than gravitational absorption, beyond what differences in spin alone would suggest.
}

\begin{document}

\maketitle

%%%%%%%%%%%%%%%%%%%%%%%%%%%%%%%%%%%%%
\section{Introduction and Summary}
\label{sec:intro}

A classical black hole absorbs all radiation incident on its horizon, making it completely opaque. Quantum fluctuations of the horizon geometry are generally expected to modify this behavior, but such effects should typically be negligible for black holes much larger than the Planck length. However, recently it has been realized that near-extremal black holes, with low temperatures and a long near-horizon throat, undergo large quantum fluctuations in the length of the throat even when the black hole is macroscopic \cite{Iliesiu:2020qvm}. Remarkably, these fluctuations are analytically tractable, as they are governed by a relatively simple quantum mechanical theory known as the Schwarzian action \cite{Mertens:2022irh}.

The impact of these quantum fluctuations on the emission and absorption of radiation has been explored in recent work \cite{Bai:2023hpd, Brown:2024ajk,Maulik:2025hax,Emparan:2025sao,Biggs:2025nzs,Lin:2025wof,Li:2025vcm}.\footnote{Radiation within the 2D throat theory was studied in \cite{Mertens:2019bvy,Blommaert:2020yeo}. Additional work on quantum corrections to the thermodynamics of non-supersymmetric systems near extremality has recently appeared in \cite{Iliesiu:2022onk,Kapec:2023ruw,Rakic:2023vhv,Emparan:2023ypa,Maulik:2024dwq,Kapec:2024zdj,Ferko:2024uxi,Kolanowski:2024zrq,Arnaudo:2024bbd,Blacker:2025zca,Mariani:2025hee,Arnaudo:2025btb}.} In particular, \cite{Emparan:2025sao,Biggs:2025nzs,Lin:2025wof} have shown that these fluctuations make the black hole appear larger when probed by low-frequency radiation of a massless scalar field: as the black hole approaches extremality, the quantum absorption cross-section for s-waves becomes increasingly greater than the semiclassical prediction. Quantum fluctuations thus enhance the observability of the black hole through such probes.

In this article, we study the absorption of electromagnetic and gravitational waves—the two massless radiation fields known to exist in nature—by a near-extremal Reissner-Nordström black hole. We find that, at sufficiently low frequencies, the absorption cross-section for a coherent, circularly polarized partial wave vanishes and the black hole becomes transparent.\footnote{If the wave is unpolarized, or linearly polarized, it can be absorbed in the form of spin-singlet di-quanta. However, this absorption is a strongly suppressed higher-order phenomenon \cite{Brown:2024ajk}.} This is in sharp contrast with the classical view of black holes as perfect absorbers: quantum effects render them invisible to light and gravitational waves below a certain frequency threshold.

\begin{figure}[t!]
    \centering
    \includegraphics[width=0.475\textwidth]{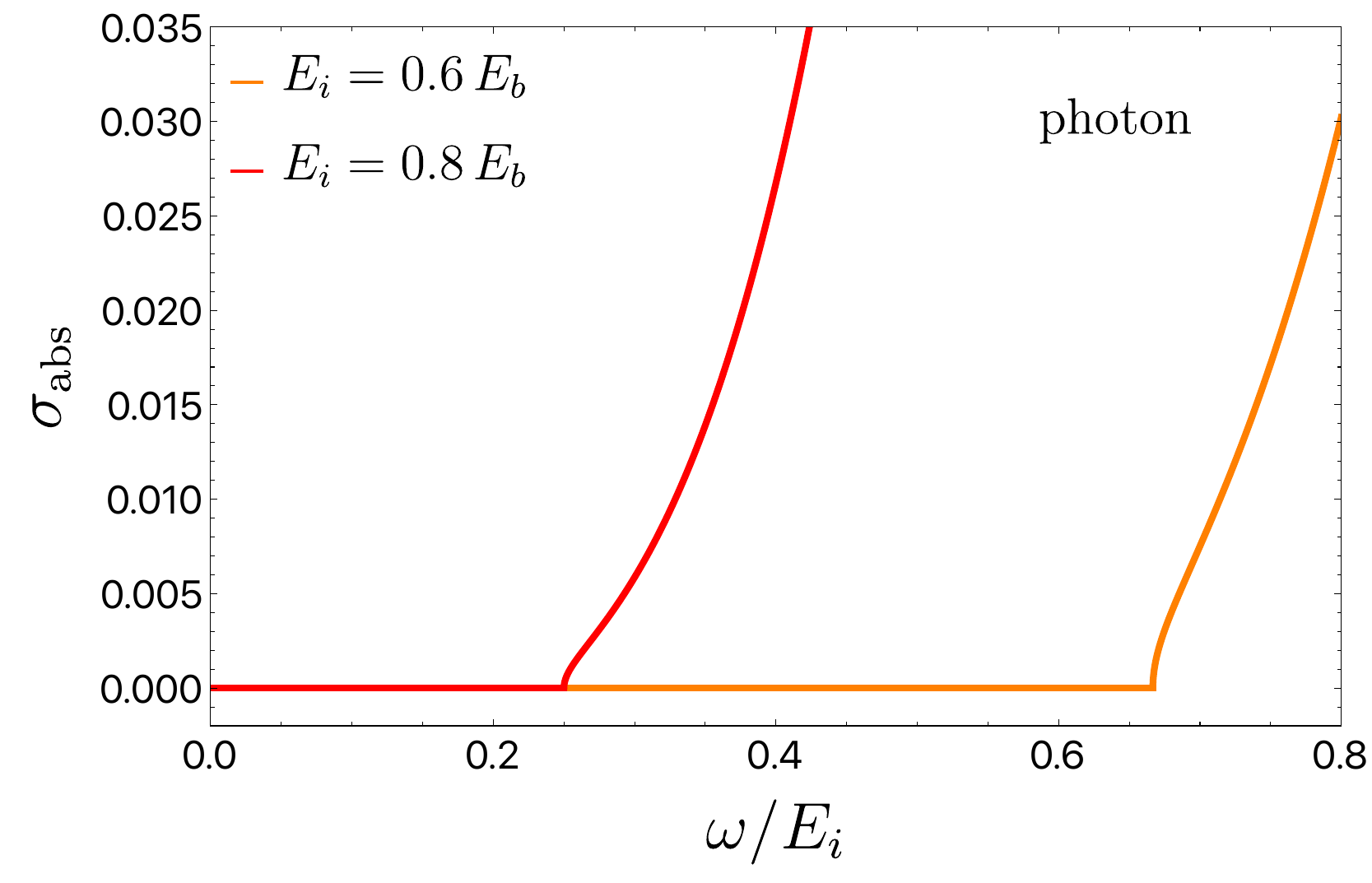}\qquad\includegraphics[width=0.475\textwidth]{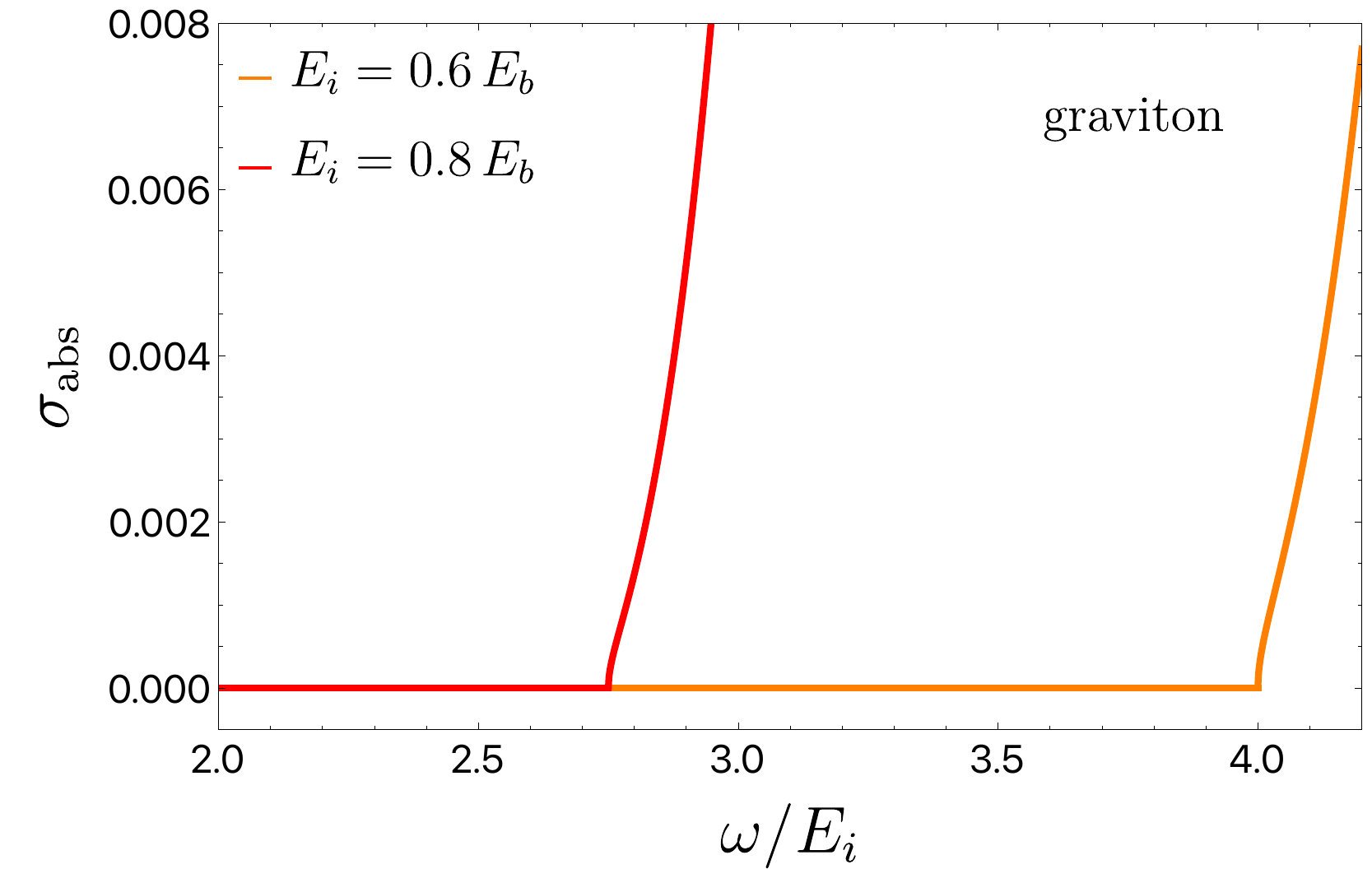}
    \caption{\small \textbf{Electromagnetic and gravitational quantum transparency}. We show the quantum absorption cross-section of photon waves with $\ell=1$ (left) and graviton waves with $\ell=2$ (right) as a function of their frequency $\omega$, for two different values of the initial black hole energy above extremality, $E_i = 0.6 \, E_b$ (orange) and $E_i = 0.8 \, E_b$ (red) (see eqs.~\eqref{eq:l1Ph} and \eqref{eq:l2Gr}). The transparency window \eqref{gapell} for $E_i = 0.6 \, E_b$ extends up to $\omega/E_i \approx 0.67$ for photons and $\omega/E_i = 4$ for gravitons. For $E_i = 0.8 \, E_b$, it extends to $\omega/E_i = 0.25$ for photons and $\omega/E_i = 2.75$ for gravitons. For the comparison between photon and graviton absorption, see Fig.~\ref{fig:l1Ph_vs_l2Gr}. The quantum and semiclassical predictions are compared in Figs.~\ref{fig:l1Ph_qu_vs_sc} and \ref{fig:l1Ph_f_rho}. In all these figures, the cross-sections are expressed in units of $E_b^{6}\, r_0^{8} = (\pi/S_0)^{6} r_0^2$, which is much smaller than the horizon area $\propto r_0^2$. This reflects the general suppression of low-frequency absorption for spinning fields, an effect that is further accentuated in the quantum regime by the strong depletion of available black hole states.}
    \label{fig:l1Ph}
\end{figure}

\paragraph{Quantized spin states of the black hole and transparency windows.} This result can be understood from basic properties of near-extremal Reissner-Nordström black holes. Quantum gravity effects become significant when the energy above extremality approaches the scale 
\begin{equation}
E_b =\frac{G}{r_0^3}\,,
\end{equation}
where $r_0 = \sqrt{G} \, Q$ is the black hole horizon radius at extremality. 
Adding quantized spin $j$ to the black hole introduces rotational energy, raising the mass of the lowest-energy state to
\begin{align}
    M_{0,j}&=M_0+\frac{G j(j+1)}{2r_0^3}\nonumber\\
    \label{eq:M0j}
    &=M_0+\frac{j(j+1)}{2}E_b\,,
\end{align}
where $M_0 = r_0 / G = Q / \sqrt{G}$ is the mass of the extremal static black hole \cite{Iliesiu:2020qvm,Heydeman:2020hhw}.\footnote{In the limit $j\gg 1$ this reproduces the energy of a slowly-rotating Kerr–Newman black hole.} The spectrum of spinning black hole states is therefore gapped above the spinless state, with a gap of order $E_b$.
%There are no black hole states with spin $j$ and mass below $M_0(j)$. 
Absorption of spin is only possible if the incident radiation carries enough energy to bridge this gap. Specifically, if a spinless black hole has initial energy $E_i=M-M_0$ above extremality, absorption of a wave with angular momentum $\ell$ requires a frequency satisfying
\begin{equation}\label{gapell}
    \omega > \frac{\ell(\ell+1)}{2}E_b-E_i\,.
\end{equation}
Thus, when we shine photons with $\ell \geq 1$ on a black hole in the quantum regime $E_i \lesssim E_b$, the black hole becomes transparent to those with frequencies $\omega < E_b - E_i$. For gravitons with $\ell \geq 2$, the transparency window extends even further, up to $\omega < 3E_b - E_i$. This phenomenon is illustrated in Figure~\ref{fig:l1Ph}.\footnote{The spin gap also constrains black hole radiation by enforcing decay selection rules \cite{Brown:2024ajk}.}

For frequencies above the bound \eqref{gapell}, the absorption of spinning waves becomes nonzero but remains small, as the density of accessible black hole states is still low. This happens despite the fact that, in the quantum regime, transition rates between individual states are enhanced, reflecting the general feature that late-time correlations in a quantum black hole decay as a power law, rather than exponentially as in the semiclassical case.  However, unlike the s-wave case studied in \cite{Emparan:2025sao}, here this enhancement is outweighed by the scarcity of absorbing black hole states across the entire frequency range. The black hole, though not entirely transparent to spinning radiation, remains very faint---i.e., translucent---and does not recover its opacity until the absorbed energy becomes large enough to push it out of the quantum regime. Among other consequences, transparency and translucency should affect the phenomenon of black-hole-induced decoherence \cite{Danielson:2022tdw,Biggs:2024dgp,Li:2025vcm}.

Additionally, our analysis also shows that electromagnetic absorption is significantly stronger than gravitational absorption, far exceeding what might be expected from differences in spin alone (Fig.~\ref{fig:l1Ph_vs_l2Gr}).

\paragraph{Generality of spin transparency.} In this article we focus on charged black holes in four-dimensional Einstein-Maxwell theory, but the results extend naturally to higher-dimensional charged black holes. A similar transparency to photons and gravitons is also expected in near-extremal neutral rotating black holes, such as the Kerr solution.\footnote{Unfortunately, the prospects of observing this effect in astrophysical black holes are vanishingly small. The scale of the relevant wavelengths is $\sim r_0 S_0$, which for a solar-mass black hole is about $10^{80}$~m, or $10^{54}$ times the size of the observable universe. Any such observation would require $\sim 10^{64}$ years, which is impractical.} 

An important difference arises in near-BPS black holes, where light fermionic modes in the throat significantly alter the dynamics \cite{Heydeman:2020hhw}. For scalar field probes, this has been studied in \cite{Lin:2025wof}, where it was found that the black hole becomes transparent to s-waves below a certain frequency, again due to a spectral gap. The gap in the near-BPS case is somewhat similar to the one above, being determined by the highest-spin component in the $j=1/2$ supermultiplet (and proportional to $j^2$ rather than $j(j+1)$), but is also present for the spin-zero excitations in the supermultiplet. It may be regarded as an enhanced level repulsion---a feature specific to systems with highly degenerate supersymmetric ground states. The spin gap examined in this article is, in this respect, a more universal feature.

We also expect analogous transparency windows for the absorption of any conserved charge whose quantization leads to similar discrete gaps in the black hole energy levels.

\paragraph{Outline of the article.} In Section~\ref{sec:sec_2} we introduce the setup. The core computations are carried out in Section~\ref{sec:sec_3}, and their physical implications are discussed in Section~\ref{sec:sec_4}. We conclude briefly in Section~\ref{sec:sec_5}. The appendices collect several lengthy technical results used throughout the paper.

%%%%%%%%%%%%%%%%%%%%%%%%%%%%%%%%%%%%%
\section{Setup}
\label{sec:sec_2}

We follow the framework used in \cite{Emparan:2025sao} for the absorption of waves by a four-dimensional near-extremal Reissner-Nordström black hole, now extended to spinning waves with $\ell \geq 1$. We will borrow results about the interaction of the black hole with various fields from \cite{Brown:2024ajk}, which studied it with a view to spontaneous Hawking emission. The technical extension to absorption is simple, but examining the response to incoming radiation places the problem in a different physical context, which brings out new aspects of quantum black hole physics.

The spinning waves we study are:
\begin{itemize}
    \item the electromagnetic field;
    \item the gravitational field;
    \item the higher partial waves of a minimally coupled massless scalar field;
    \item a minimally coupled massless vector, of a $U(1)$ gauge field different from the one the black hole is charged under---a `dark' photon.
\end{itemize}

\subsection{Scattering waves from infinity into the throat} 

We consider a monochromatic, circularly polarized wave of these fields with frequency $\omega$ and angular momentum $\ell$, which we model as a coherent state of the field with a large occupation number. This wave propagates classically on the background geometry from asymptotic infinity down to the mouth of the black hole throat, where it encounters the black hole as a quantum object. The semiclassical approximation breaks down there, and the propagation of the wave in the throat geometry is replaced by the interaction between the wave and the quantum states of the black hole described by the Schwarzian theory. This interaction accounts for the quantum backreaction of the black hole to the incoming radiation.

In a bit more detail (for further discussion, see \cite{Brown:2024ajk,Emparan:2025sao}), when the field $\Phi(t,r,\theta,\varphi)$ 
is expanded in angular modes $\Phi_{\ell m}(t,r)$ and propagated to the mouth of the throat,  it acts there as a source 
that sets a boundary condition for the effective two-dimensional field $\phi_\Delta(t,r)$ in AdS$_2$. According to standard AdS/CFT reasoning, the value of the field near the AdS$_2$ boundary as $z \to 0$ (with $z \sim 1/r$)
\begin{equation}
    \phi_\Delta(t,z) \sim \phi_\text{bdy}(t) \, z^{1-\Delta} + \dots
\end{equation}
couples to a dual conformal operator $\mathcal{O}_\Delta(t)$ which describes the response of the black hole. The conformal dimension $\Delta$ depends on $\ell$ and the type of field under consideration, and is the only data needed for the coupling to the throat theory---we will give the specific values shortly. Then, the system of the quantum black hole and its coupling to the field is captured by the action
\begin{equation}
    \label{eq:total_action}
    I = I_\text{Schw} + \int dt \, \phi_\text{bdy}(t) \, \mathcal{O}_\Delta(t) \, ,
\end{equation}
where $I_\text{Schw}$ is the Schwarzian action.

This general framework describes the interaction between the black hole and an external field in an arbitrary state, but our focus is on the case where the field consists of incoming and outgoing waves. Expanding the field near asymptotic infinity in terms of ingoing-mode components $A_{\omega \ell m}$, their propagation to the mouth of the throat leads to\footnote{We are taking $A_{\omega \ell m}$ to have dimensions of inverse-square-root of time, so that, as we will see, its amplitude squared gives the ingoing particle flux.}
\begin{equation}
    \label{eq:phi_bdy_sM_ell}
    \phi_\text{bdy}(t) = \mathcal{N}_\ell \,\, r_0^{\Delta + \ell}  \omega^{\ell + 1/2} \, \left(A_{\omega \ell m} \, e^{-i \omega t} + \text{h.c.} \right) \, .
\end{equation}
The factor $\omega^{\ell + 1/2}$ reflects the short-distance suppression of spinning waves as they tunnel through the angular momentum barrier at low frequencies. The factor $r_0^{\Delta + \ell}$ sets the correct dimensionful scale, ensuring that the normalization constant $\mathcal{N}_\ell$ remains a dimensionless, $\ell$-dependent number.\footnote{From the perspective of the far zone, the black hole is indistinguishable from a classical extremal solution characterized solely by the scale $r_0$. The temperature scale enters only through subleading corrections.} In principle, $\mathcal{N}_\ell$ can be derived by carefully normalizing the quantum theory in the throat region and matching it to the far-zone solution. However, a more practical route is to determine it by requiring that the quantum absorption probabilities reduce to the correct leading low-frequency value in the semiclassical limit for each field (see Appendix~\ref{app:A}). In this sense, the reader may view these normalizations as having been inferred, in reverse, from our results below for the full quantum cross-section, as was indeed done in \cite{Brown:2024ajk,Emparan:2025sao}.

For the fields that we consider the results are 
\begin{itemize}
    \item Minimal massless scalar with $\ell \ge 0$:
     \begin{equation}
    \label{eq:DeMl_scalar}
    \Delta=\ell+1\,,\qquad |\mathcal{N}_\ell|^2 = \frac{\Gamma(\ell + 1)^2}{4\pi \, \Gamma(2 \ell + 1) \, \Gamma\!\left(\ell + \frac{3}{2} \right)^2} \, .
    \end{equation}
        \item Photon with $\ell = 1$:
    \begin{equation}
    \label{eq:DeMl_l1_photon}
        \Delta = \ell + 2 = 3 \, , \qquad\qquad |\mathcal{N}_1|^2 = \frac{40}{9 \pi^2} \, .
    \end{equation}
    \item Photon with $\ell \ge 2$:
    \begin{equation}
    \label{eq:DeMl_lge2_photon}
        \Delta = \ell\,,\qquad |\mathcal{N}_\ell|^2 = \frac{(\ell - 1) \, (\ell + 1)^2 \, \Gamma(\ell - 1)^2}{16 \pi \, \Gamma(2 \ell) \, \Gamma\!\left(\ell + \frac{3}{2} \right)^2} \, .
    \end{equation}
    \item Graviton with $\ell \ge 2$:
    \begin{equation}
    \label{eq:DeMl_graviton}
        \Delta = \ell\,,\qquad |\mathcal{N}_\ell|^2 = \frac{(\ell + 2) \, (\ell + 1)^2 \, \Gamma(\ell - 1)^2}{16 \pi \, \Gamma(2 \ell) \, \Gamma\!\left(\ell + \frac{3}{2} \right)^2} \, .
    \end{equation}
    \item Dark photon with $\ell \ge 1$:
    \begin{equation}
    \label{eq:DeMl_hidden_photon}
        \Delta = \ell + 1\,,\qquad |\mathcal{N}_\ell|^2 = \frac{2^{4 \ell} \, \Gamma(\ell)^2 \, \Gamma(\ell + 2)^2}{\pi^2 \, \Gamma(2 \ell + 1) \, \Gamma(2 \ell + 2)^2} \, .
    \end{equation}
\end{itemize}

Observe that the $\ell = 1$ photon values \eqref{eq:DeMl_l1_photon} are not the $\ell \to 1$ limit of those for $\ell \geq 2$ \eqref{eq:DeMl_lge2_photon}. This is because, in the vicinity of a charged black hole, graviton modes mix with photon modes for $\ell \geq 2$. This electromagnetic-gravitational mixing alters the propagation of the fields, affecting both their absorption and reflection coefficients \cite{Crispino:2009zza,Oliveira:2011zz,Brown:2024ajk}.

An interesting consequence is that in the throat theory photons and gravitons with the same $\ell \geq 2$ are dual to operators with the same scaling dimension $\Delta$, since they effectively correspond to the same modes in this region. Their propagation outside the throat depends only on $\ell$, implying that their absorption and emission rates exhibit the same frequency dependence and differ solely by an overall factor of $(\ell + 2)/(\ell - 1)$, as noted in \cite{Brown:2024ajk}. We will return to this point in Section~\ref{subsec:phvsgr}.

The case of the dark photon is different. Although its absorption coefficients match those of massless scalar waves with $\ell \geq 1$, up to an $\ell$-dependent factor, this is not the result of a symmetry or coupling. Unlike the electromagnetic photon, the dark photon does not mix with other fields and remains fully decoupled from them. As a consequence, the throat conformal dimensions of these two photons are different.

\subsection{Transition rates}

At this stage, computing the transition rates between the initial and final states of the black hole-radiation system becomes a time-dependent perturbation problem in quantum mechanics, with a perturbation potential
\begin{equation}
    \label{eq:Vt_V0}
    V(t) = \left(V_0 \, e^{-i \omega t} + \text{h.c.} \right) \, \mathcal{O}_\Delta(t) \, , \qquad V_0 = -\mathcal{N}_\ell \,\, r_0^{\Delta + \ell} \, \omega^{\ell + 1/2} \, A_{\omega \ell m} \, .
\end{equation}
When the perturbation is weak we can apply Fermi's Golden Rule, which gives us the absorption and stimulated emission transition rates in the form\footnote{To apply Fermi's Golden Rule, the interaction between the incoming wave and the quantum Schwarzian must last a time $t$ such that $1/(E_i \pm \omega) \ll t \ll 1/| \langle f | V_0 \, \mathcal{O}_\Delta | i \rangle |$, where the plus sign is for absorption and the minus sign for stimulated emission. The first bound guarantees that only the final states with $E_f \approx E_i \pm \omega$ contribute to the transition probability. The second bound ensures that the transition probability always remains small (i.e., that our perturbation remains weak). For the parameter range we are interested in, such a time window always exists, and it is typically extremely large, thanks to the exponential $e^{-S_0}$ in the matrix elements \eqref{eq:2pt_O}.} 
\begin{align}
    \label{eq:Ga_abs_1}
    \Gamma_\text{abs} &= \left.2 \pi \, | \langle f | V_0 \, \mathcal{O}_\Delta | i \rangle |^2 \, \rho(E_f) \right|_{E_f = E_i+\omega} \, ,\\
    \label{eq:Ga_stem_1}
    \Gamma_\text{em} &= \left.2 \pi \, | \langle f | V_0 \, \mathcal{O}_\Delta | i \rangle |^2 \, \rho(E_f) \right|_{E_f = E_i-\omega} \, ,
\end{align}
where $|i\rangle$ and $|f\rangle$ are the initial and final states, with energies $E_i$ and $E_f$ (which are always measured above extremality), $\mathcal{O}_\Delta \equiv \mathcal{O}_\Delta(0)$, and $\rho(E_f)$ is the density of final states.\footnote{Note that, although the black hole starts in a fixed-energy state, it will evolve through Hawking emission into an ensemble of states in a time $\sim E_b^{-1}$, comparable to the duration of our scattering experiment \cite{Biggs:2025nzs}.} As in \cite{Emparan:2025sao}, we can treat the radiation states as eigenstates of the occupation number %$N_\omega$ 
instead of coherent states, since for large occupation numbers their variance is negligible. For the same reason, we can neglect spontaneous emission, i.e., Hawking radiation.

Let us now account also for the angular momentum of the radiation and the black hole.
We use $j$ and $j'$ for the initial and final spin of the black hole, and $m_\text{\tiny BH}$ and $m'_\text{\tiny BH}$ for the initial and final azimuthal mode numbers. Each quantum of the incoming wave has energy $\omega$ and angular momentum numbers $\ell$ and  $m$. Note that it must be $j' \subset j \otimes \ell$, so we have 
\begin{equation}
    |j-\ell| \leq j' \leq j+\ell\,, \qquad m'_\text{\tiny BH} = m_\text{\tiny BH} + m\,.
\end{equation}
Then \eqref{eq:Ga_abs_1} and \eqref{eq:Ga_stem_1} become
\begin{align}
    \label{eq:Ga_abs_2}
    \Gamma_\text{abs} &= 2 \pi \, |\mathcal{N}_\ell|^2 \, \mathcal{F}_\omega \, r_0^{2 (\Delta + \ell) } \, \omega^{2 \ell + 1} \, | \langle E_i + \omega, j', m'_\text{\tiny BH} | \mathcal{O}_\Delta | E_i, j, m_\text{\tiny BH} \rangle |^2 \, \rho_{j'}(E_i + \omega) \, ,\\
    \label{eq:Ga_stem_2}
    \Gamma_\text{em} &= 2 \pi \, |\mathcal{N}_\ell|^2 \, \mathcal{F}_\omega \, r_0^{2 (\Delta + \ell) }\, \omega^{2 \ell + 1} \, | \langle E_i - \omega, j', m'_\text{\tiny BH} | \mathcal{O}_\Delta | E_i, j, m_\text{\tiny BH} \rangle |^2 \, \rho_{j'}(E_i - \omega) \, \Theta(E_i - \omega) \, ,
\end{align}
where we inserted the expression of $V_0$ in \eqref{eq:Vt_V0} and $|\langle A_{\omega\ell m}\rangle|^2 =\mathcal{F}_\omega$ is the incoming flux.
The step function $\Theta(E_i - \omega)$ indicates that there cannot be any emission for $\omega > E_i$. Our notation makes it explicit that the density of final states $\rho$ depends on their spin $j'$.

To compute the total absorption cross-section for a given partial wave of angular momentum $\ell$ with fixed circular polarization, we must average over the initial angular momentum states $m_\text{\tiny BH}$, and sum over the possible values of the final angular momentum $j'$ of the black hole, as well as over the quantum numbers $m'_\text{\tiny BH}$ and $m$.

Henceforth we will restrict our attention to the case of a spinless initial black hole, $j=0$. We do this for simplicity and because our interest lies in the comparison between the quantum answer and the semiclassical one, which is $j$-independent. However, the analysis can be readily extended to initially spinning black holes. 

With this choice, the absorption of the wave will result in a spinning black hole with
\begin{equation}
    j' = \ell\,.
\end{equation}

The  optical theorem gives the total absorption cross-section for a single polarization mode of the incoming wave
\begin{equation}
    \label{eq:sigmaabs_Pgb}
    \sigma_\text{abs} = \frac{(2 \ell + 1) \, \pi}{\omega^2}P_\text{gb}
\end{equation}
in terms of the greybody factor
\begin{equation}\label{Pgb}
    P_\text{gb} = \frac{2 \pi}{\mathcal{F}_\omega} \, \delta_{j 0} \, \delta_{j'\ell} \; \sum_{m'_\text{\tiny BH}, m_\text{\tiny BH}, m} \left(\Gamma_\text{abs} - \Gamma_\text{em}\right) \, .
\end{equation}
This factor is the net absorption probability, summed over all the final quantum numbers $m'_\text{\tiny BH}$ and $m$. For an initial spinning black hole one should also average over the values of $m_\text{\tiny BH}$, but since we start with a static black hole, then in our sums $m_\text{\tiny BH}=0$. 

%%%%%%%%%%%%%%%%%%%%%%%%%%%%%%%%%%%%%
\section{Computing the cross-section}
\label{sec:sec_3}

To obtain the absorption and emission rates \eqref{eq:Ga_abs_2} and \eqref{eq:Ga_stem_2} we require the density of black hole states and the probabilities for the transitions between them induced by the operator $\mathcal{O}_\Delta$. These have been derived in the literature on the Schwarzian theory, as reviewed in \cite{Mertens:2022irh}. 

To keep expressions reasonably compact it is convenient to introduce some notation, following \cite{Brown:2024ajk,Heydeman:2020hhw}. First, recalling the expression for the mass at extremality for a black hole with angular momentum $j$, i.e., $M_{0,j}$ in \eqref{eq:M0j}, we denote the energy gap above the spinless extremal state by
\begin{equation}
    \label{eq:E0j_def}
    E_{0,j} = M_{0,j} - M_0 =\frac{j \, (j + 1)}{2} \, E_b\,.
\end{equation}
Furthermore, we define
\begin{equation}
    \label{eq:GammaDeltaEiEf}
    \Gamma^\Delta_{E_i,E_f} = \Gamma\left(\Delta \pm i \sqrt{\frac{2 E_i}{E_b}}\pm i \sqrt{\frac{2 E_f}{E_b}} \right) \, ,
\end{equation}
with the usual convention that $\Gamma^\Delta_{E_i,E_f}$ is the product of the four Gamma functions that appear for each  choice of sign.

Then, the results we need can be written as
\begin{equation}
    \label{eq:2pt_O}
    | \langle E_f, j', m'_\text{\tiny BH} | \mathcal{O}_\Delta | E_i, j, m_\text{\tiny BH} \rangle |^2 = \frac{\left|C^{j' \, m'_\text{\tiny BH}}_{j \, m_\text{\tiny BH}, \, \ell \, m} \right|^2}{2 j' + 1} \, \frac{2 \, e^{-S_0} \, \Gamma^\Delta_{E_i,E_f}}{(2/E_b)^{2 \Delta} \, \Gamma(2 \Delta)} \, ,
\end{equation}
and
\begin{equation}
    \label{eq:rho_Ef}
    \rho_{j'}(E_f) = \frac{(2 j'+ 1) \, e^{S_0}}{2 \, \pi^2 \, E_b} \, \sinh\!\left(2 \pi \, \sqrt{2 \, (E_f - E_{0,j'})/E_b} \right) \, \Theta(E_f - E_{0,j'}) \,,
\end{equation}
where $S_0$ is the semiclassical black hole entropy at extremality, and $C^{j' \, m'_\text{\tiny BH}}_{j \, m_\text{\tiny BH}, \, \ell \, m}$ are Clebsch-Gordan coefficients of $\text{SU}(2)$. Here, as in \cite{Brown:2024ajk}, we use the \textsl{Mathematica 14} conventions for them. When averaging over initial states and summing over final ones, we employ the identity
\begin{equation}
    \label{eq:CG_identity}
    \sum_{m'_\text{\tiny BH}, m_\text{\tiny BH}, m} \left|C^{j' \, m'_\text{\tiny BH}}_{j \, m_\text{\tiny BH}, \, \ell \, m} \right|^2 = 2 j' + 1\,,
\end{equation}  
bearing in mind that our initial spinless black hole has $j=0$, $m_\text{\tiny BH}=0$ and $j' = \ell$.

Now, with a little extra piece of notation,
\begin{equation}
    \label{eq:sAEf}
    \mathcal{A}^\Delta_{E_f} = \frac{\Gamma^\Delta_{E_i, E_f}}{(2 \pi)^2 \, \omega/E_b} \, \sinh\!\left(2 \pi \, \sqrt{\frac{2 \, (E_f - E_{0,\ell})}{E_b} }\right) \, \Theta(E_f - E_{0,\ell}) \, ,
\end{equation}
we can write the greybody factor as
\begin{equation}
    \label{eq:Pgb}
    P_\text{gb} = |\mathcal{N}_\ell|^2 \,\, \frac{\pi^2 \, (2 \ell + 1)}{2^{2 (\Delta - 2)} \, \Gamma(2 \Delta)} \, E_b^{2 (\Delta - 1)} \, r_0^{2 (\Delta + \ell)} \, \omega^{2 (\ell + 1)} \, \left(\mathcal{A}^\Delta_{E_i + \omega} - \mathcal{A}^\Delta_{E_i - \omega} \right) \, .
\end{equation}

Note that the step function $\Theta(E_i - \omega)$ in \eqref{eq:Ga_stem_2}---which enforces the condition that no emission occurs to states below the extremal black hole---no longer appears in \eqref{eq:Pgb}. This is because $\mathcal{A}^\Delta_{E_i - \omega}$ imposes the stricter constraint that $\omega < E_i - E_{0,\ell}$, meaning that no emission is possible to states lying below the spin-induced energy gap.

These expressions can be further simplified by noting that, for all the fields we consider, the scaling dimension $\Delta$ is a positive integer. Then we can use the Pochhammer symbols
\begin{equation}
(a)_n = a \, (a+1) \cdots (a+n-1) = \frac{\Gamma(a+n)}{\Gamma(a)}
\end{equation}
to write
\begin{equation}
    \label{eq:GammaDeltaEiEf_2}
    \Gamma^\Delta_{E_i, E_i \pm \omega} = \mathcal{B}^\Delta_{E_i \pm \omega} \, \Gamma^1_{E_i, E_i \pm \omega} \, ,
\end{equation}
where
\begin{equation}\label{Gamma1}
    \Gamma^1_{E_i, E_i \pm \omega} = \pm \frac{(2 \pi)^2 \, \omega/E_b}{\cosh\!\left(2 \pi \, \sqrt{2 (E_i \pm \omega)/E_b} \right) - \cosh\!\left(2 \pi \, \sqrt{2 E_i/E_b} \right)} \, ,
\end{equation}
and the previous Gammas have been replaced by Pochhammers in the form
\begin{align}
    \label{eq:sBp_1}
    \mathcal{B}^\Delta_{E_i + \omega} &= \left(1 \pm i \, \sqrt{2 E_i / E_b} \pm i \, \sqrt{2 (E_i + \omega)/E_b} \right)_{\Delta - 1}\\
    \label{eq:sBp_2}
    &= \left(1 \pm \sqrt{-2 \left(\omega + 2 E_i \pm 2 \sqrt{E_i (E_i + \omega)} \right)/E_b} \right)_{\Delta - 1} \, ,\\
    \label{eq:sBm_1}
    \mathcal{B}^\Delta_{E_i - \omega} &= \left(1 \pm i \, \sqrt{2 E_i / E_b} \pm i \, \sqrt{2 (E_i - \omega)/E_b} \right)_{\Delta - 1}\\
    \label{eq:sBm_2}
    &= \left(1 \pm \sqrt{2 \left(\omega - 2 E_i \pm 2 \sqrt{E_i (E_i - \omega)} \right)/E_b} \right)_{\Delta - 1} \, .
\end{align}
Here we use again the convention that denotes a product of Pochhammer symbols for the four choices of sign. The products nicely simplify to a single product of real polynomials in $E_i/E_b$ and $\omega/E_b$,
\begin{equation}
    \label{eq:sBpm_prod}
     \mathcal{B}^\Delta_{E_i \pm \omega} = \prod_{n=1}^{\Delta - 1} \! \left[n^4 + \frac{4 n^2 \, (2 E_i \pm \omega)}{E_b} + \frac{4 \omega^2}{E_b^2} \right] \, .
\end{equation}

These expressions are convenient since the conformal dimension $\Delta$ is entirely contained in the factors ${B}^\Delta_{E_i \pm \omega}$ that multiply the term \eqref{Gamma1}, which gave the transition rates in \cite{Emparan:2025sao} for $\Delta=1$. 
Now, for $E_f = E_i \pm \omega$ the expression \eqref{eq:sAEf} of $\mathcal{A}^\Delta_{E_f}$ reduces to
\begin{equation}
    \label{eq:sAEf_2}
    \mathcal{A}^\Delta_{E_i \pm \omega} = \pm \frac{\mathcal{B}^\Delta_{E_i \pm \omega} \, \sinh\!\left(2 \pi \, \sqrt{2 \, (E_i \pm \omega - E_{0,\ell}) / E_b} \right) \, \Theta(E_i \pm \omega - E_{0,\ell})}{\cosh\!\left(2 \pi \, \sqrt{2 (E_i \pm \omega)/E_b} \right) - \cosh\!\left(2 \pi \, \sqrt{2 E_i/E_b} \right)} \, .
\end{equation}

Collecting all results, the quantum absorption cross-section takes the form
\begin{equation}
    \label{eq:sigmaabs_general}
    \sigma_\text{abs} = |\mathcal{N}_\ell|^2 \, \frac{\pi^3 \, (2 \ell + 1)^2}{2^{2 (\Delta - 2)} \, \Gamma(2 \Delta)} \, E_b^{2 (\Delta - 1)} \, r_0^{2 (\Delta + \ell)} \, \omega^{2 \ell} \, \left(\mathcal{A}^\Delta_{E_i + \omega} - \mathcal{A}^\Delta_{E_i - \omega} \right) \, ,
\end{equation}
where $\mathcal{A}^\Delta_{E_i \pm \omega}$ is given by \eqref{eq:sAEf_2}, and $|\mathcal{N}_\ell|^2$ is defined in \eqref{eq:DeMl_scalar}--\eqref{eq:DeMl_hidden_photon}. This constitutes our main result.

Since this general expression is somewhat unwieldy, in the following section we give the explicit form of the cross-section for the $\ell = 1$ photon and $\ell=2$ graviton in \eqref{eq:l1Ph} and \eqref{eq:l2Gr}, while the specific expressions for the other types of fields we consider are presented in Appendix~\ref{app:C}.

\paragraph{Semiclassical limit.} We can now derive the semiclassical limit $\omega, E_b \ll E_i$ of the quantum greybody factor \eqref{eq:Pgb}. In this regime, the Pochhammers \eqref{eq:sBpm_prod} reduce to
\begin{align}
    \mathcal{B}^\Delta_{E_i \pm \omega} &\approx \prod_{n=1}^{\Delta - 1} \! \left(\frac{8 n^2  E_i}{E_b} + \frac{4 \omega^2}{E_b^2} \right)\\
    &= \left(\!\frac{4 \pi}{E_b \beta} \!\right)^{2 (\Delta - 1)} \, \left(\!1 + i \, \frac{\beta \omega}{2 \pi} \!\right)_{\Delta - 1} \, \left(\!1 - i \, \frac{\beta \omega}{2 \pi} \!\right)_{\Delta - 1}\\
    \label{eq:sB_sc}
    &= \frac{E_b}{2 \pi \omega} \, \left(\!\frac{4 \pi}{E_b \beta} \!\right)^{2 \Delta - 1} \, \sinh\!\left(\!\frac{\beta \omega}{2} \!\right) \, \Gamma\!\left(\!\Delta + i \, \frac{\beta \omega}{2 \pi} \!\right) \, \Gamma\!\left(\!\Delta - i \, \frac{\beta \omega}{2 \pi} \!\right) \, ,
\end{align}
where the inverse temperature near extremality is given by
\begin{equation}
     \label{eq:T_leading}
     \beta = \frac{2 \pi}{\sqrt{2 E_b \, E_i}}\, .
\end{equation}
% where we have used the identity
% \begin{equation}
%     \sinh\!\left(\!\frac{\beta \omega}{2} \!\right) \, \Gamma\!\left(\!\Delta \pm i \, \frac{\beta \omega}{2 \pi} \!\right) = \frac{2 \pi^2}{\beta \omega} \, \left(i \, \frac{\beta \omega}{2 \pi} \right)_\Delta \, \left(-i \, \frac{\beta \omega}{2 \pi} \right)_\Delta
% \end{equation}
% to express the Pochhammer symbols in \eqref{eq:Pgb} in terms of Gamma functions. 

In this limit we have
\begin{equation}
    \label{eq:Th_sc}
    \Theta(E_i \pm \omega - E_{0,\ell}) \to 1 \, .
\end{equation}
This helps simplify the term $\mathcal{A}^\Delta_{E_i + \omega} - \mathcal{A}^\Delta_{E_i - \omega}$ in the cross-section. Up to the Pochhammer factors \eqref{eq:sB_sc}, it is now proportional to
\begin{align}
    &\frac{\sinh\!\left(2 \pi \, \sqrt{\frac{2 \, (E_i + \omega - E_{0,\ell})}{E_b}} \right)}{\cosh\!\left(2 \pi \, \sqrt{\frac{2 (E_i + \omega)}{E_b}} \right) - \cosh\!\left(2 \pi \, \sqrt{\frac{2 E_i}{E_b}} \right)} - \frac{\sinh\!\left(2 \pi \, \sqrt{\frac{2 \, (E_i - \omega - E_{0,\ell})}{E_b}} \right)}{\cosh\!\left(2 \pi \, \sqrt{\frac{2 E_i}{E_b}} \right) - \cosh\!\left(2 \pi \, \sqrt{\frac{2 (E_i - \omega)}{E_b}} \right)} \nonumber\\
    &\approx e^{-\beta E_{0,\ell}} \, \left(\frac{e^{\beta \omega}}{e^{\beta \omega} - 1} - \frac{e^{-\beta \omega}}{1 - e^{-\beta \omega}} \right) = e^{-\beta E_{0,\ell}} \approx 1 \, .
\end{align}
Then $\mathcal{A}^\Delta_{E_i + \omega} - \mathcal{A}^\Delta_{E_i - \omega}$ is given by \eqref{eq:sB_sc} and the semiclassical greybody factor reduces to
\begin{equation}
    \label{eq:PgbSC}
    P_\text{gb} \approx |\mathcal{N}_\ell|^2 \,\, \frac{2 \, (2 \pi)^{2 \Delta} \, (2 \ell + 1)}{\Gamma(2 \Delta)} \, \frac{r_0^{2 (\Delta + \ell)} \, \omega^{2 \ell + 1}}{\beta^{2 \Delta - 1}} \, \sinh\!\left(\!\frac{\beta \omega}{2} \!\right) \, \Gamma\!\left(\!\Delta \pm i \, \frac{\beta \omega}{2 \pi} \!\right)\,.
\end{equation}

In Appendix~\ref{app:A} we show that when the constants $|\mathcal{N}_\ell|^2$ are chosen as in \eqref{eq:DeMl_scalar}--\eqref{eq:DeMl_hidden_photon}, the expression \eqref{eq:PgbSC} reproduces, in unified form, all the semiclassical greybody factors previously derived in \cite{Crispino:2000jx,Page:2000dk,Crispino:2009zza,Oliveira:2011zz,Arbey:2021jif, Arbey:2021yke} from wave propagation across the full frequency range $\omega\ll 1/r_0$ (including $\beta \omega =\mathcal{O}(1)$). 

This demonstrates that the semiclassical results emerge as a limit of the full quantum expression \eqref{eq:Pgb}. Remarkably, \eqref{eq:PgbSC} captures both cases with and without electromagnetic-gravitational mixing, unifying all the semiclassical results within a single compact formula.

It is also illustrative to take the very low-frequency limit $\omega \ll \beta^{-1}$. Our result \eqref{eq:PgbSC} simplifies to 
\begin{equation}
    \label{eq:Pgb_qu_sc}
    P_\text{gb} \approx |\mathcal{N}_\ell|^2 \,\, \frac{\pi^2 \, (2 \ell + 1) \, 2^{\Delta + 1} \, \Gamma(\Delta)^2}{\Gamma(2 \Delta)} \, (r_0 \, E_b)^{2 (\Delta + \ell)} \, \left(\!\frac{E_i}{E_b} \!\right)^{\Delta + 2 \ell + 1} \, \left(\!\frac{\omega}{E_i} \!\right)^{2 (\ell + 1)} \, .
\end{equation}
This shows the parametric dependence of the absorption probability more clearly, and makes it slightly easier to determine the constants $|\mathcal{N}_\ell|^2$ from the comparison to the limit $\omega \ll \beta^{-1}$ of the semiclassical wave-propagation results in Appendix~\ref{app:A}.

%%%%%%%%%%%%%%%%%%%%%%%%%%%%%%%%%%%%%
\section{Analysis of results}
\label{sec:sec_4}

\subsection{Quantum spin transparency}
From \eqref{eq:sAEf_2} and \eqref{eq:sigmaabs_general}, we see that the necessary conditions to have absorption and (stimulated) emission are
\begin{align}
    \label{eq:abs_cond}
    \omega &> E_{0,\ell} - E_i \qquad \text{for absorption},\\
    \label{eq:stem_cond}
    \omega &< E_i - E_{0,\ell} \qquad \text{for emission} \, 
\end{align}
 (cf.~\eqref{gapell}). 
 
 Let us first discuss the case where the initial energy of the spinless black hole is \emph{above} the spin gap of the final spinning black hole, i.e., 
 \begin{equation}
     E_i > E_{0,\ell}\,.
 \end{equation}
 In this case there is no obstacle to absorption for any frequency $\omega > 0$, while stimulated emission will occur only if it takes the black hole down to a state above the spin gap, i.e., $\omega < E_i - E_{0,\ell}$. When no spin is involved, i.e., $\ell=0$, stimulated emission is restricted to $\omega < E_i$, which leads to the kinks in Figure 1 of \cite{Emparan:2025sao} for the absorption cross-section as a function of $\omega$.\footnote{We take this opportunity to correct a typo in Figs.~2 and 3 of \cite{Emparan:2025sao}: the caption states $E_i = 0.1\, E_b$, but the plots actually correspond to $E_i = 0.01\, E_b$.}

If instead the initial energy is \emph{below} the spin-gap of the final black hole,
\begin{equation}
    E_i < E_{0,\ell}\,,
\end{equation}
which includes the regime $E_i \lesssim E_b$ where a genuinely quantum Schwarzian description is needed, then no emission can occur, and absorption is possible only if \eqref{eq:abs_cond} is satisfied. In other words, the black hole is transparent when \begin{equation}
\label{eq:transp_cond}
    \omega < E_{0,\ell} - E_i \qquad \text{for transparency}.
\end{equation}

In the analysis and figures below, we focus on this regime with $\ell \ge 1$ and $E_i < E_{0,\ell}$, where the distinctive quantum features of transparency and absence of stimulated emission emerge. 

The spin-induced gap also closes the channel for single-quantum emission, suppressing spontaneous emission. This led \cite{Brown:2024ajk} to consider spin-singlet pairs of entangled photons or gravitons, whose spontaneous emission typically produces a slight difference between ingoing and outgoing fluxes. In our setup, however, the fixed-helicity radiation state has zero overlap with such di-photon or di-graviton states, so there is no absorption or stimulated emission of these di-quanta within the window \eqref{eq:transp_cond}, and the absorption cross-section vanishes exactly.

Interestingly, a generic coherent wave---such as an unpolarized electromagnetic or gravitational wave with support across multiple angular momenta $\ell$---will generally have nonzero overlap with di-quanta states. The same applies to linearly polarized waves. In such cases, absorption and stimulated emission of di-quanta occur, and the black hole is no longer perfectly transparent. These second-order processes are, however, strongly suppressed \cite{Brown:2024ajk}, so the black hole remains much fainter in the window \eqref{eq:transp_cond} than in the regime \eqref{eq:abs_cond} above the spin-induced gap.

Throughout, we plot the absorption cross-section $\sigma_\text{abs}$  for fixed values of $E_i/E_b$ as a function of the rescaled frequency $\omega/E_i$. The natural unit of length that we use for the cross-section is $E_b^{\Delta + \ell - 1}\,  r_0^{\Delta + \ell}$. Since $E_b =\pi/(r_0 S_0)$, this becomes $(\pi/S_0)^{\Delta+\ell-1} r_0$. For all fields except s-wave massless scalars ($\Delta = 1$, $\ell = 0$), this scale is much smaller than $r_0$, reflecting the suppression of absorption at low frequencies due to spin or conformal dimension. As we will show in Section~\ref{subsec:qvssc}, this suppression is reinforced in the quantum regime by the drastic reduction in the number of available states.

\subsection{Results for $\ell=1$ photon and $\ell=2$ graviton}\label{subsec:phvsgr}

We focus on the $\ell = 1$ photon and the $\ell = 2$ graviton modes, which dominate the absorption of electromagnetic and gravitational radiation by the black hole. 
Their total absorption cross-sections are given respectively by
\begin{align}
    \label{eq:l1Ph}
    &\sigma_\text{abs} = \frac{\pi}{3} \, r_0^8 \, \omega^2 \, \times\nonumber\\
    \Bigg[&\frac{\left(4 E_b^2 + 4 E_b (2 E_i + \omega) + \omega^2 \right) \left(E_b^2 + 4 E_b (2 E_i + \omega) + 4 \omega^2 \right) \sinh\!\left(\!2 \pi \sqrt{\frac{2 (E_i + \omega - E_b)}{E_b}} \right) \Theta(E_i + \omega - E_b)}{\cosh\!\left(\!2 \pi \sqrt{\frac{2 (E_i + \omega)}{E_b}} \right) - \cosh\!\left(\!2 \pi \sqrt{\frac{2 E_i}{E_b}} \right)} \nonumber\\
    &-\frac{\left(4 E_b^2 + 4 E_b (2 E_i - \omega) + \omega^2 \right) \left(E_b^2 + 4 E_b (2 E_i - \omega) + 4 \omega^2 \right) \sinh\!\left(\!2 \pi \sqrt{\frac{2 (E_i - \omega - E_b)}{E_b}} \right) \Theta(E_i - \omega - E_b)}{\cosh\!\left(\!2 \pi \sqrt{\frac{2 E_i}{E_b}} \right) - \cosh\!\left(\!2 \pi \sqrt{\frac{2 (E_i - \omega)}{E_b}} \right)} \Bigg]
\end{align}
and
\begin{align}
    \label{eq:l2Gr}
    \sigma_\text{abs} =  &\,\frac{4 \pi}{9} \, r_0^8 \, \omega^4 \, \times\nonumber\\
    \Bigg[&\frac{\left(E_b^2 + 4 E_b (2 E_i + \omega) + 4 \omega^2 \right) \sinh\!\left(\!2 \pi \sqrt{\frac{2 (E_i + \omega - 3 E_b)}{E_b}} \right) \Theta(E_i + \omega - 3 E_b)}{\cosh\!\left(\!2 \pi \sqrt{\frac{2 (E_i + \omega)}{E_b}} \right) - \cosh\!\left(\!2 \pi \sqrt{\frac{2 E_i}{E_b}} \right)} \nonumber\\
    &-\frac{\left(E_b^2 + 4 E_b (2 E_i - \omega) + 4 \omega^2 \right) \sinh\!\left(\!2 \pi \sqrt{\frac{2 (E_i - \omega - 3 E_b)}{E_b}} \right) \Theta(E_i - \omega - 3 E_b)}{\cosh\!\left(\!2 \pi \sqrt{\frac{2 E_i}{E_b}} \right) - \cosh\!\left(\!2 \pi \sqrt{\frac{2 (E_i - \omega)}{E_b}} \right)} \Bigg] \, .
\end{align}
These are shown in Figure~\ref{fig:l1Ph}. Both these cross-sections display the transparency window \eqref{eq:transp_cond}. 
Above this window, the cross-sections do not exhibit any kinks since, as we explained, stimulated emission is completely absent.

\begin{figure}[t!]
    \centering
    \includegraphics[width=0.65\textwidth]{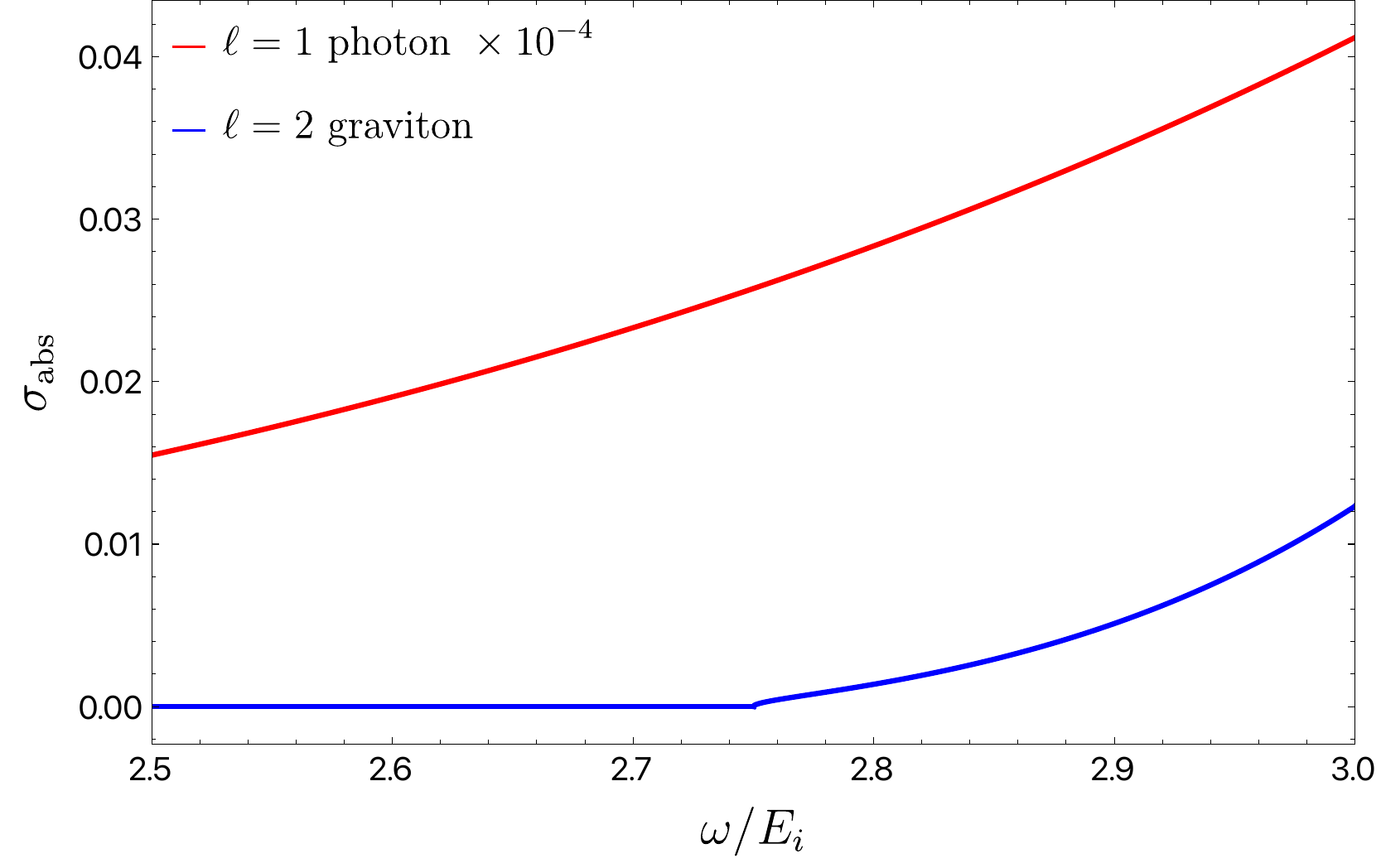}
    \caption{\small Quantum absorption cross-section of $\ell = 1$ photon \eqref{eq:l1Ph} (red) vs.\ $\ell = 2$ graviton \eqref{eq:l2Gr} (blue) as a function of $\omega$ for a black hole with $E_i = 0.8 \, E_b$. The photon cross-section has been reduced by a factor of $10^{-4}$ to bring it within the figure---see the text for the explanation of the dramatic enhancement of the photon absorption. The transparency window \eqref{eq:transp_cond} for the photon ends at $\omega / E_i = 0.25$ (not visible in the figure), while for the graviton it ends at $\omega/E_i = 2.75$.}
    \label{fig:l1Ph_vs_l2Gr}
\end{figure}

Comparing the two, the cross-section for the photon dominates, as shown in Figure~\ref{fig:l1Ph_vs_l2Gr}. 
As a result, the $\ell = 1$ photon provides the leading absorption channel in Einstein gravity coupled to the Standard Model.

At first sight, this may seem natural since the $\ell=1$ photon faces a lower angular momentum barrier than the $\ell = 2$ graviton. However, the difference is surprisingly large: to visually compare them in Figure~\ref{fig:l1Ph_vs_l2Gr}, we had to rescale the photon cross-section by a factor of $10^{-4}$. 

This dramatic enhancement originates from the electromagnetic–gravitational mixing present in the charged black hole background. As a result, the conformal dimension of the operator dual to the $\ell=2$ graviton is $\Delta=2$, which is smaller than that of the $\ell=1$ photon, $\Delta = 3$. % larger than any of the naively expected values $\Delta=1$ or possibly $\Delta=2$ (see eqs.~\eqref{eq:DeMl_l1_photon}, \eqref{eq:DeMl_lge2_photon}, and \eqref{eq:DeMl_hidden_photon}). 

\begin{figure}[t!]
    \centering
    \includegraphics[width=0.475\textwidth]{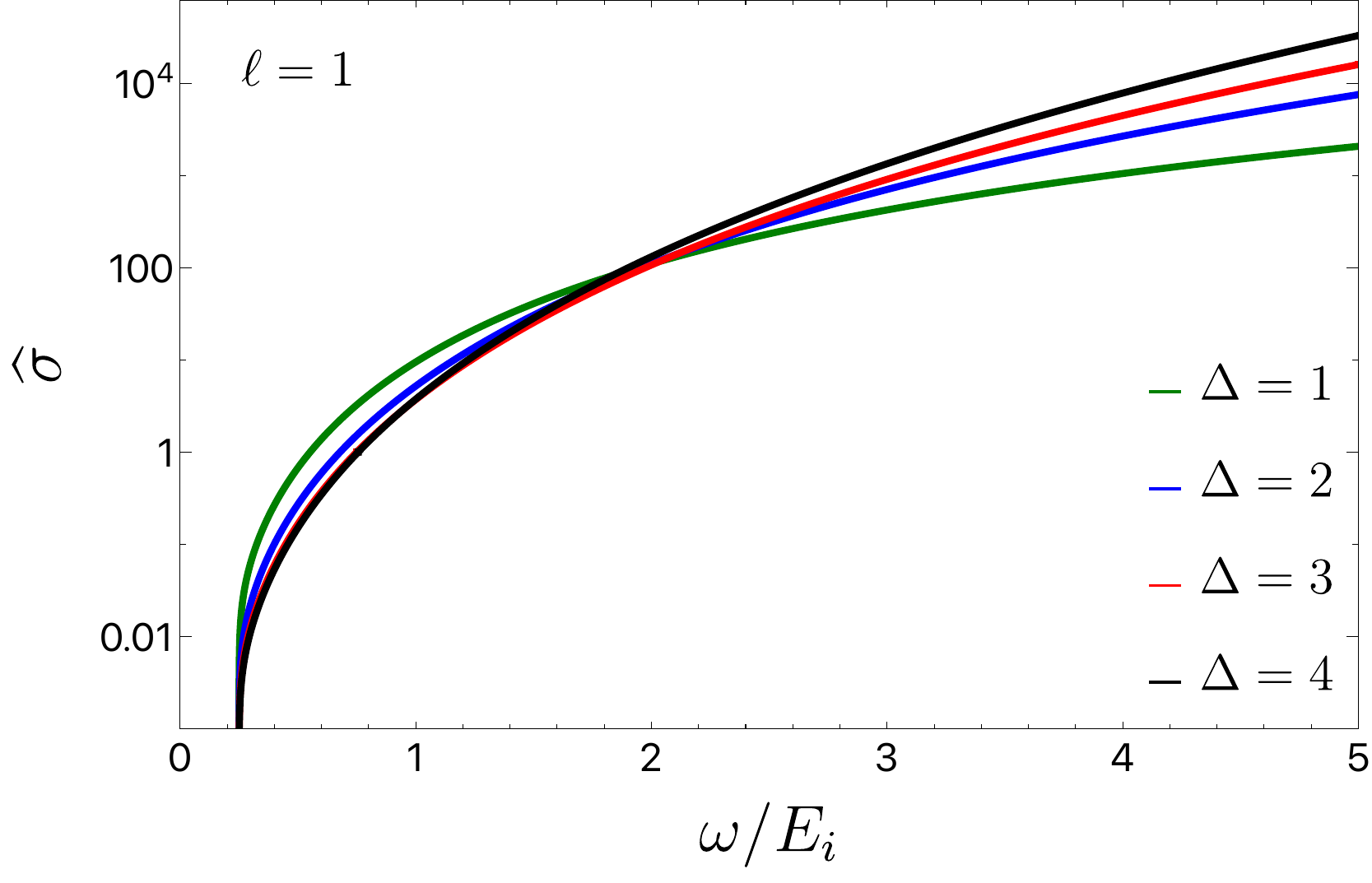}\qquad\includegraphics[width=0.475\textwidth]{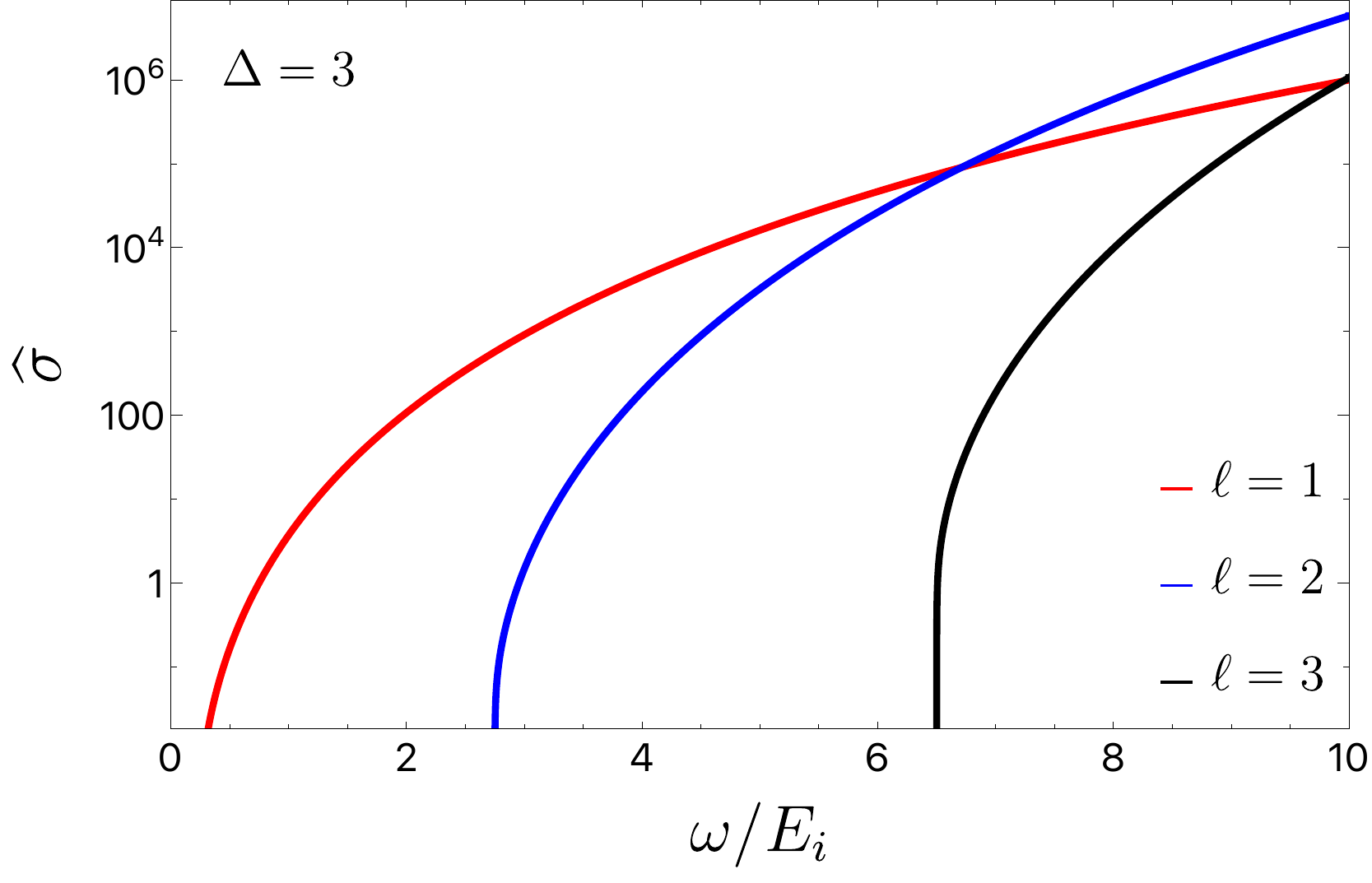}
    \caption{\small Field-type-independent absorption cross-section $\widehat{\sigma}$ (cf.~\eqref{eq:sigmahat}) as a function of $\omega$ for $E_i = 0.8 \, E_b$. Left: plot at fixed $\ell = 1$ for different $\Delta$. Right:  plot at fixed $\Delta = 3$ for different $\ell$. In the quantum regime where $\omega/E_i\lesssim \mathcal{O}(1)$ (say, $\omega/E_i \le 5$), but above the transparency window for the $\ell = 2$ graviton, which ends at $\omega/E_i = 2.75$, the cross-section becomes higher as we increase $\Delta$, and as we decrease $\ell$. The same qualitative feature is observed for fixed $\ell = 2$, or for fixed $\Delta = 2$. The length unit for $\widehat{\sigma}$ is $E_b^{\Delta+\ell-1}\, r_0^{\Delta+\ell}$.}
    \label{fig:sigmahat}
\end{figure}

Let us dig a bit more on this point. As a general rule, since $\Delta$ increases with $\ell$, it tends to suppress absorption, at least in semiclassical frequency regimes. However, this holds only within a given type of field, and as we have seen, the relationship between $\ell$ and $\Delta$ varies across different fields. To disentangle their individual effects on absorption, we remove the field-dependent normalization factor $\mathcal{N}_\ell$,
\begin{equation}
\label{eq:sigmahat}
    \widehat{\sigma} = \frac{\sigma_\text{abs}}{|\mathcal{N}_\ell|^2}\, ,
\end{equation}
so $\widehat{\sigma}$ is identical for any two fields with the same $\ell$ and $\Delta$. In the frequency range of interest,\footnote{Namely, above the $\ell=2$ graviton transparency window and within the quantum regime. For the value of $E_i$ that we are considering in Figs.~\ref{fig:l1Ph_vs_l2Gr} and \ref{fig:sigmahat}, namely, $E_i = 0.8 \, E_b$, this range is $2.75 \leq \omega/E_i \lesssim 5$.} $\widehat{\sigma}$ shows a clear pattern: it decreases with increasing $\ell$ and increases with increasing $\Delta$. Figure~\ref{fig:sigmahat} illustrates this trend for fixed $\ell = 1$ (left panel) and fixed $\Delta = 3$ (right panel); the same qualitative behavior is observed for fixed $\ell = 2$ and fixed $\Delta = 2$. Since the density of states is independent of $\Delta$, the observed enhancement in $\widehat{\sigma}$ with increasing $\Delta$ arises from a corresponding increase in the transition probabilities induced by $\mathcal{O}_\Delta$ in the frequency range where we make the comparison.

The type of field also affects the absorption cross-section through a different mechanism than the precise relation between $\Delta$ and $\ell$, namely, the normalization factor $\mathcal{N}_\ell$. As discussed in Section~\ref{sec:sec_2}, the $\ell = 2$ graviton cross-section is four times larger than that of the $\ell = 2$ photon, even though both share the same values of $\Delta$ and $\ell$ and hence have the same frequency dependence. The disparity arises from their normalization constants, which differ by a factor of $(\ell + 2)/(\ell - 1) = 4$.

\subsection{Absorption of radiation with large frequency $\omega\gg E_i$}

It is worth observing that our framework remains valid even when studying the absorption of radiation with frequencies much larger than the initial energy, $\omega \gg E_i$. Although such high-frequency radiation pushes the black hole out of the quantum regime, it does not necessarily disrupt its long near-extremal throat. 

The most complete disruption would occur if the final black hole energy $E_f$ were large enough that the corresponding inverse temperature, $\beta_f \sim 1/\sqrt{E_b E_f}$, which sets the throat length, became comparable to the throat thickness $r_0$. For $\omega \gg E_i$ we have $E_f \sim \omega$ and the throat remains parametrically long compared to its width as long as
\begin{equation}\label{upperbound}
    \omega \ll \frac{1}{E_b  r_0^2}\sim M_0\, .
\end{equation}
where we used $E_b\sim 1/(r_0 S_0)$, and $M_0 \sim S_0/r_0$ is the black hole mass near extremality. 

This result is expected: a significant modification of the geometry requires an energy injection comparable to the black hole mass. However, this bound is too loose. When studying semiclassical wave propagation in the far zone, we assume a stricter condition,
\begin{equation}\label{upperbound2}
    \omega \ll 1/r_0 \sim S_0 E_b\,.
\end{equation}
This ensures that the fractional change in the mass of the black hole, as perceived by distant observers, remains small $\lesssim 1/S_0$. Nevertheless, it still permits energy absorption far exceeding $E_b$. Accordingly, our framework remains applicable for black holes initially in the quantum regime ($E_i \lesssim E_b$) that absorb large energies $\omega \gg E_i$, as long as the bound \eqref{upperbound2} is respected.

\subsection{Quantum vs.~semiclassical absorption: emergence of quantum translucency}
\label{subsec:qvssc}

\begin{figure}[t!]
    \centering
    \includegraphics[width=0.65\textwidth]{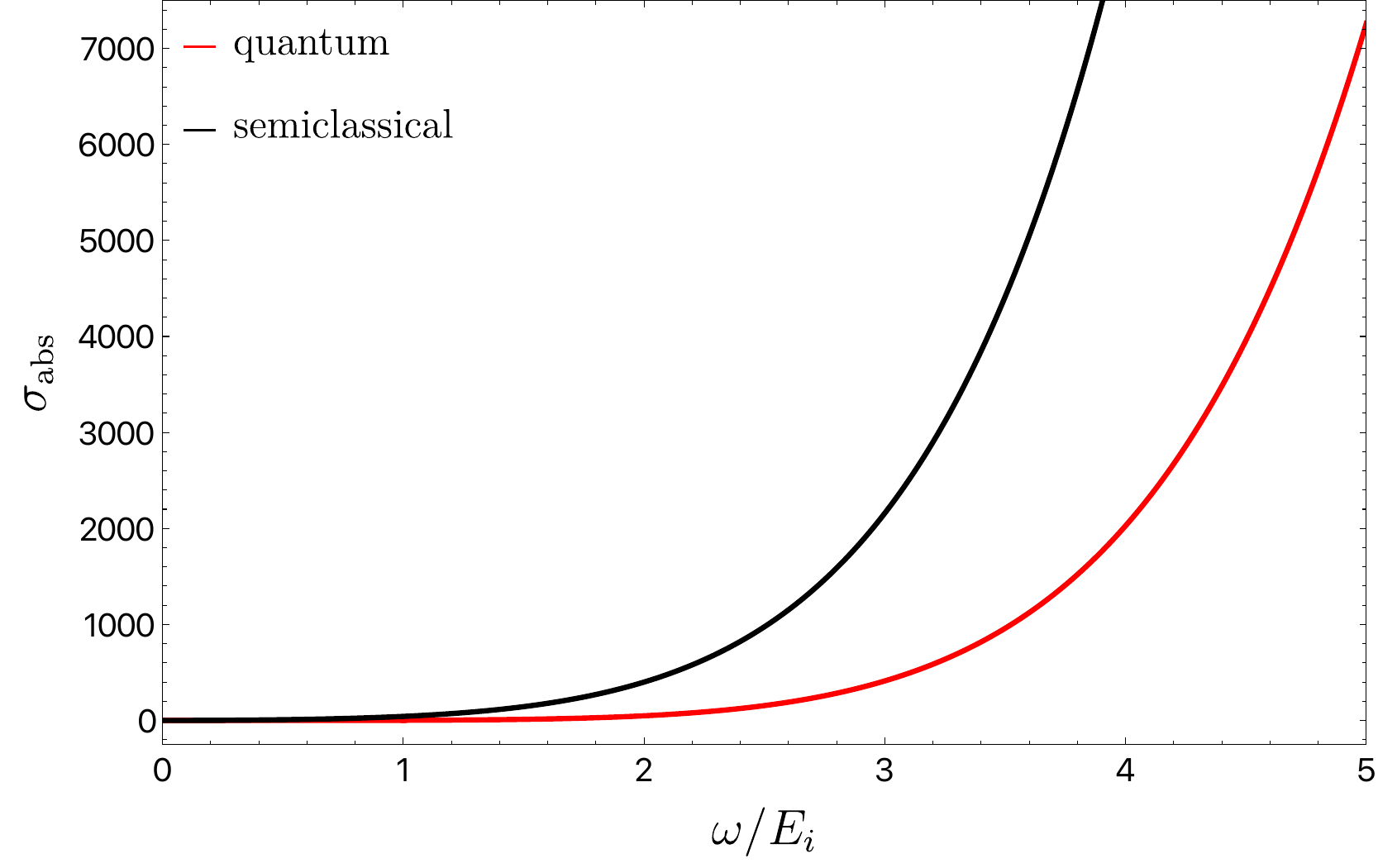}
    \caption{\small Quantum \eqref{eq:l1Ph} (red) vs.\ semiclassical \eqref{eq:Pgb_full_sc_l1_photon} (black) absorption cross-section of the $\ell = 1$ photon as a function of $\omega$, for $E_i = 0.8 \, E_b$. The quantum transparency window \eqref{eq:transp_cond} ends at $\omega / E_i = 0.25$ and is not easily distinguishable. Note how the quantum cross-section is always suppressed compared to the semiclassical one, giving rise to quantum translucency.}
    \label{fig:l1Ph_qu_vs_sc}
\end{figure}

As seen in Figure~\ref{fig:l1Ph}, for both the $\ell = 1$ photon and the $\ell = 2$ graviton, increasing $E_i/E_b$---that is, approaching the semiclassical black hole regime---leads to a larger absorption cross-section. This is in sharp contrast with the behavior of the $\ell = 0$ scalar studied in \cite{Emparan:2025sao}, where the cross-section decreases as $E_i/E_b$ increases. The underlying reason is that, for $\ell \geq 1$, the semiclassical absorption cross-section dominates over the quantum one, whereas the opposite holds for $\ell = 0$. To illustrate this difference, Figure~\ref{fig:l1Ph_qu_vs_sc} compares the quantum and semiclassical results for the $\ell = 1$ photon.\footnote{When $\omega\gg E_i$ (which is \emph{not} the semiclassical limit) the quantum result approaches the semiclassical value, since $E_f \approx \omega$ is dominated by the strong backreaction of the incoming wave, blurring the differences between the quantum initial state and its semiclassical limit.}

The salient feature is that above the transparency window, the quantum cross-section becomes nonzero but remains below the semiclassical one---the black hole is neither transparent nor opaque, but translucent. As we will see, this is due to the still low density of available final states.

To understand this effect, we analyze separately the contributions of the density of final states, $\rho_\ell(E_f)$, and the transition probabilities
\begin{equation}
    \label{eq:f_def}
    f_\Delta(E_f) = \delta_{j 0} \, \delta_{j'\ell} \; \sum_{m'_\text{\tiny BH}, m_\text{\tiny BH}, m} | \langle E_f, j', m'_\text{\tiny BH} | \mathcal{O}_\Delta | E_i, j, m_\text{\tiny BH} \rangle |^2\,,
\end{equation}
to the total absorption cross-section $\sigma_\text{abs}$ (technical details can be found in Appendix~\ref{app:B}). 

In Figure~\ref{fig:l1Ph_f_rho} we compare the quantum expressions for the transition probabilities and density of states with their semiclassical counterparts (i.e., \eqref{eq:f_qu} and \eqref{eq:rho_qu} vs.\ \eqref{eq:f_sc} and \eqref{eq:rho_sc}), in the quantum regime $E_i\lesssim E_b$ (where, recall, there is no stimulated emission of spin).
\begin{figure}[t!]
    \centering
    \includegraphics[width=0.475\textwidth]{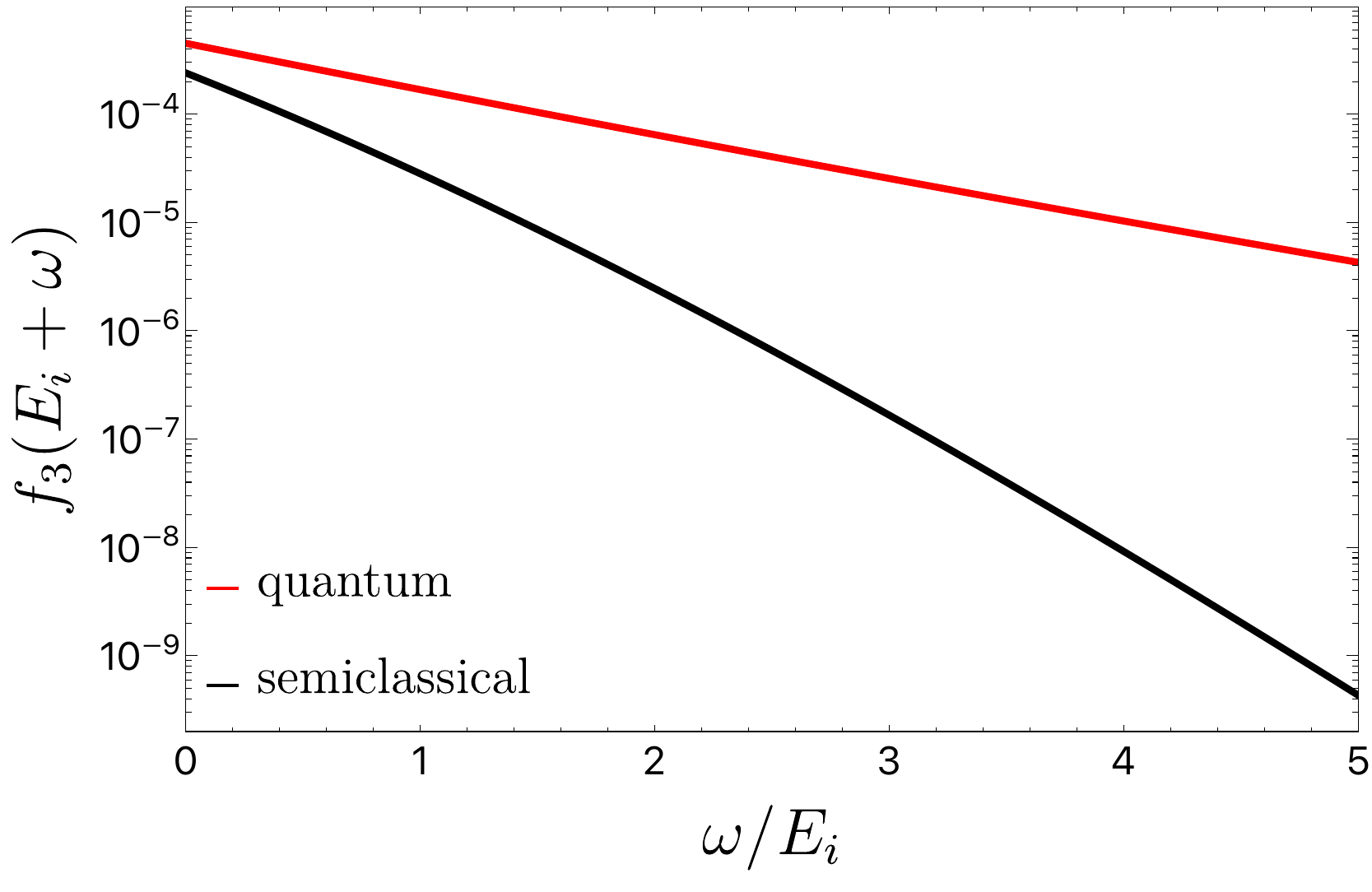}\qquad\includegraphics[width=0.475\textwidth]{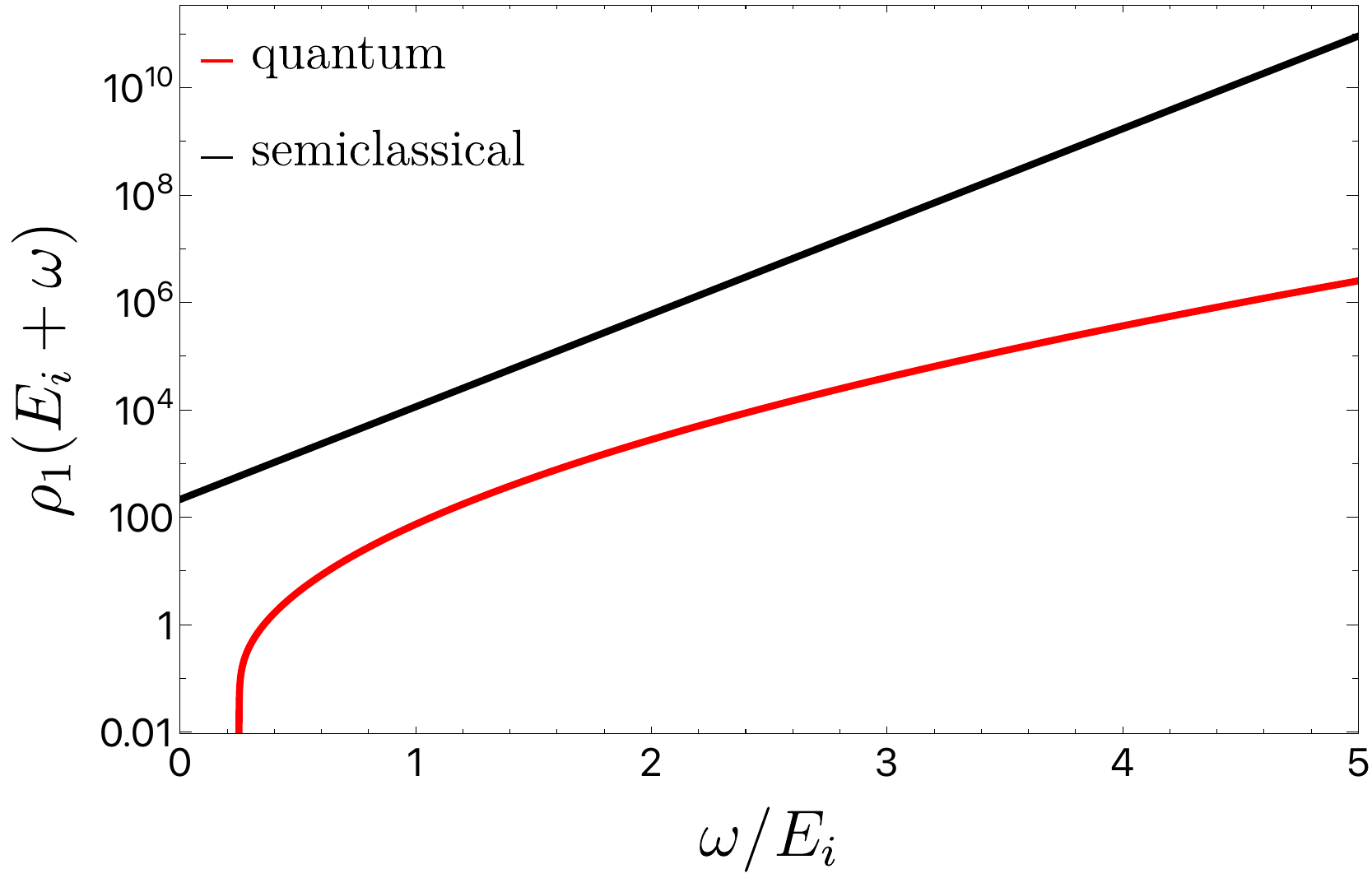}
    \caption{\small Quantum vs.\ semiclassical absorption transition probabilities (left) and density of final states (right) of the $\ell = 1$ photon as a function of $\omega$, for $E_i = 0.8 \, E_b$ (eqs.~\eqref{eq:f_qu}--\eqref{eq:rho_sc}). The transparency window \eqref{eq:transp_cond} ends at $\omega / E_i = 0.25$. As in the case of the $\ell = 0$ scalar \cite{Emparan:2025sao}, quantum fluctuations markedly enhance absorption transitions while sharply suppressing the density of final states. These competing effects determine whether the absorption cross-section is enhanced or suppressed, the latter being the case for spinning radiation. We plot $f_3$ in units of $e^{-S_0} E_b^6$, and $\rho_1$ in units of $e^{S_0} E_b^{-1}$. The log-scale vertical axes highlight the large disparity between semiclassical and quantum values.}
    \label{fig:l1Ph_f_rho}
\end{figure}
This comparison reveals the origin of the semiclassical dominance: while quantum states absorb spinning radiation more efficiently (left panel), their scarcity relative to semiclassical expectations (right panel) results in a lower overall quantum cross-section---in contrast to the case of scalar s-waves studied in \cite{Emparan:2025sao,Biggs:2025nzs}, where quantum effects actually enhanced absorption. In the present case, quantum absorption remains suppressed throughout the entire range \eqref{upperbound2} where our expressions are valid, creating the phenomenon of quantum translucency.

%%%%%%%%%%%%%%%%%%%%%%%%%%%%%%%%%%%%%
\section{Closing remarks}\label{sec:sec_5}

Hawking famously showed that quantum mechanics makes black holes less black by enabling them to emit radiation. Our results reveal that quantum gravity effects can go even further: at sufficiently low temperatures, charged black holes become transparent to low-frequency radiation, failing to absorb it altogether. 

The existence of these transparency windows highlights the fact that black holes are genuine quantum systems. Although most of the black hole spectrum is extremely dense---with level spacings of order $\Delta E \sim e^{-S}$, which appear continuous in semiclassical gravity---quantized spin induces much larger energy gaps, scaling as $\Delta E \sim 1/S^\#$. Remarkably, this regime is accessible via gravitational path integral methods, offering a consistent and powerful approach to probe the quantum black hole spectrum beyond the semiclassical approximation.

%%%%%%%%%%%%%%%%%%%%%%%%%%%%%%%%%%%%%
\section*{Acknowledgments}

Work supported by MICINN grant PID2022-136224NB-C22, AGAUR grant 2021 SGR 00872, and grant CEX2024-001451-M funded by MICIU/AEI/10.13039/501100011033. The project that gave rise to these results received the support of a fellowship from ``la Caixa” Foundation (ID 100010434), awarded to ST, with code LCF/BQ/DI24/12070002.

%%%%%%%%%%%%%%%%%%%%%%%%%%%%%%%%%%%%%
\appendix
\section{Semiclassical greybody factors}
\label{app:A}

The semiclassical greybody factors for the fields considered in this work have been computed in \cite{Crispino:2000jx,Page:2000dk,Crispino:2009zza,Oliveira:2011zz,Arbey:2021jif, Arbey:2021yke}, and we use them in the form presented in \cite{Bai:2023hpd,Brown:2024ajk}. To fix the normalization factor $\mathcal{N}_\ell$ in the quantum absorption formulas, it is sufficient to consider the leading-order behavior in the low-frequency limit $\omega \ll \beta^{-1}$, where near extremality $\beta$ is given by \eqref{eq:T_leading}.

The expressions from \cite{Bai:2023hpd,Brown:2024ajk} and their low-frequency limits are:
\begin{itemize}
    \item Minimal massless scalar with $\ell \ge 0$:
    \begin{equation}
        \label{eq:Pgb_full_sc_scalar}
        P_\text{gb} = \frac{2^{3 (2 \ell + 1)} \, \pi^{2 \ell} \, \Gamma(\ell + 1)^4}{\Gamma(2 \ell + 1)^2 \, \Gamma(2 \ell + 2)^2} \, \left(\!\frac{r_0^2 \, \omega}{\beta} \!\right)^{2 \ell + 1} \, \sinh\!\left(\!\frac{\beta \omega}{2} \!\right) \, \Gamma\!\left(\!\ell + 1 \pm i \, \frac{\beta \omega}{2 \pi} \!\right) \, .
    \end{equation}
    %\begin{equation}
    %\label{eq:Pgb_full_sc_scalar}
    %    P_\text{gb} = \frac{\pi^{2 (\ell + 2)} \, r_0^{2 (2 \ell + 1)} \, (\omega / \beta)^{2 \ell + 1} \, \text{csch}(\beta \omega / 2)}{2^{2 \ell - 1} \, \Gamma\!\left(\ell + \frac{1}{2} \right)^2 \, \Gamma\!\left(\ell + \frac{3}{2} \right)^2 \, \Gamma\!\left(\!-\ell \pm i \, \frac{\beta \omega}{2 \pi} \!\right)} \, .
    %\end{equation}
    \\
    Low-frequency limit:
    \begin{equation}
    \label{eq:Pgb_sc_scalar}
        P_\text{gb} \approx \frac{\pi^2 \, \Gamma(\ell + 1)^2}{2^{3 \ell} \, \Gamma\!\left(\ell + \frac{1}{2} \right)^2 \, \Gamma\!\left(\ell + \frac{3}{2} \right)^2} \, (r_0 \, E_b)^{2 (2 \ell + 1)} \, \left(\!\frac{E_i}{E_b} \!\right)^{3 \ell + 2} \, \left(\!\frac{\omega}{E_i} \!\right)^{2 (\ell + 1)} \, .
    \end{equation}
    \item Photon  with $\ell = 1$:
    \begin{equation}
    \label{eq:Pgb_full_sc_l1_photon}
        P_\text{gb} = \frac{4}{9} \, \left(r_0^2 \, \omega \right)^4 \, \left(\frac{4 \, \pi^2}{\beta^2} + \omega^2 \right) \, \left(\frac{16 \, \pi^2}{\beta^2} + \omega^2 \right) \, .
    \end{equation}
    \\
    Low-frequency limit:
    \begin{equation}
    \label{eq:Pgb_sc_l1_photon}
        P_\text{gb} \approx \frac{64}{9} \, (r_0 \, E_b)^8 \, \left(\! \frac{E_i}{E_b} \!\right)^6 \, \left(\! \frac{\omega}{E_i} \!\right)^4 \, .
    \end{equation}
    \item Photon  with $\ell \ge 2$:
    \begin{equation}
    \label{eq:Pgb_full_sc_lge2_photon}
        P_\text{gb} = (\ell - 1) \, \frac{\pi^{2 \ell} \, (\ell + 1)^2}{2^{2 \ell - 1} \, (2 \ell + 1) \, (\ell - 1)^2 \, \Gamma\!\left(\ell + \frac{1}{2} \right)^4} \, \frac{r_0^{4 \ell} \, \omega^{2 \ell + 1}}{\beta^{2 \ell - 1}} \, \sinh\!\left(\!\frac{\beta \omega}{2} \!\right) \, \Gamma\!\left(\!\ell \pm i \, \frac{\beta \omega}{2 \pi} \!\right) \, .
    \end{equation}
    \\
    Low-frequency limit:
    \begin{equation}
    \label{eq:Pgb_sc_lge2_photon}
        P_\text{gb} \approx (\ell - 1) \, \frac{\pi^2 \, (\ell + 1)^2 \, \Gamma(\ell - 1)^2}{2^{3 \ell - 1} \, (2 \ell + 1) \, \Gamma\!\left(\ell + \frac{1}{2} \right)^4} \, (r_0 \, E_b)^{4 \ell} \, \left(\!\frac{E_i}{E_b} \!\right)^{3 \ell + 1} \, \left(\!\frac{\omega}{E_i} \!\right)^{2 (\ell + 1)} \, .
    \end{equation}
    \item Graviton  with $\ell \ge 2$:
    \begin{equation}
    \label{eq:Pgb_full_sc_graviton}
        P_\text{gb} = (\ell + 2) \, \frac{\pi^{2 \ell} \, (\ell + 1)^2}{2^{2 \ell - 1} \, (2 \ell + 1) \, (\ell - 1)^2 \, \Gamma\!\left(\ell + \frac{1}{2} \right)^4} \, \frac{r_0^{4 \ell} \, \omega^{2 \ell + 1}}{\beta^{2 \ell - 1}} \, \sinh\!\left(\!\frac{\beta \omega}{2} \!\right) \, \Gamma\!\left(\!\ell \pm i \, \frac{\beta \omega}{2 \pi} \!\right) \, .
    \end{equation}
    \\
    Low-frequency limit:
    \begin{equation}
    \label{eq:Pgb_sc_graviton}
        P_\text{gb} \approx (\ell + 2) \, \frac{\pi^2 \, (\ell + 1)^2 \, \Gamma(\ell - 1)^2}{2^{3 \ell - 1} \, (2 \ell + 1) \, \Gamma\!\left(\ell + \frac{1}{2} \right)^4} \, (r_0 \, E_b)^{4 \ell} \, \left(\!\frac{E_i}{E_b} \!\right)^{3 \ell + 1} \, \left(\!\frac{\omega}{E_i} \!\right)^{2 (\ell + 1)} \, .
    \end{equation}
    % Note that, for $\ell \ge 2$, the ratio between the graviton and the photon greybody factors is simply $(\ell + 2)/(\ell - 1)$.
    \item Dark photon  with $\ell \ge 1$:
    \begin{equation}
    \label{eq:Pgb_full_sc_hidden_photon}
        P_\text{gb} = \frac{2^{3 (2 \ell + 1)} \, \pi^{2 \ell} \, \Gamma(\ell)^2 \, \Gamma(\ell + 2)^2}{\Gamma(2 \ell + 1)^2 \, \Gamma(2 \ell + 2)^2} \, \left(\!\frac{r_0^2 \, \omega}{\beta} \!\right)^{2 \ell + 1} \, \sinh\!\left(\!\frac{\beta \omega}{2} \!\right) \, \Gamma\!\left(\!\ell + 1 \pm i \, \frac{\beta \omega}{2 \pi} \!\right)\,.
    \end{equation}
    \\
    Low-frequency limit:
    \begin{equation}
    \label{eq:Pgb_sc_hidden_photon}
        P_\text{gb} \approx \frac{2^{\ell + 2} \, \pi \, \Gamma(\ell)^2 \, \Gamma(\ell + 2)^2}{\Gamma\!\left(2 \ell + 2 \right)^2 \, \Gamma\!\left(\ell + \frac{1}{2} \right)^2} \, (r_0 \, E_b)^{2 (2 \ell + 1)} \, \left(\!\frac{E_i}{E_b} \!\right)^{3 \ell + 2} \, \left(\!\frac{\omega}{E_i} \!\right)^{2 (\ell + 1)} \, .
    \end{equation}
\end{itemize}
Comparing these results with the semiclassical, low-frequency limit \eqref{eq:Pgb_qu_sc} of the quantum greybody factor yields the normalization factors $\mathcal{N}_\ell$ for the various types of fields.

Furthermore, it is straightforward to verify that inserting into the semiclassical limit \eqref{eq:PgbSC} of the quantum greybody factor the appropriate values of $|\mathcal{N}_\ell|^2$ and $\Delta$ from \eqref{eq:DeMl_scalar}--\eqref{eq:DeMl_hidden_photon}, we recover the complete semiclassical greybody factors given above for all fields.

\section{Explicit absorption cross-sections}
\label{app:C}

In this appendix, we present the explicit expressions for the quantum absorption cross-sections of the field types considered in the main text. The functions $\mathcal{B}^{\Delta}_{E_i \pm \omega}$ are defined in \eqref{eq:sBp_1}--\eqref{eq:sBpm_prod}, and should be evaluated using the appropriate values of $\Delta$ for each field, as specified in \eqref{eq:DeMl_scalar}--\eqref{eq:DeMl_hidden_photon}.

\begin{itemize}
    \item Minimal massless scalar with $\ell \ge 0$:
    \begin{equation}
        \label{eq:sigmaabs_scalar}
        \sigma_\text{abs} = \frac{\pi^3}{(2 \ell + 1) \, 2^{2 (3 \ell - 1)} \, \Gamma\!\left(\ell + \frac{1}{2} \right)^4} \, E_b^{2 \ell} \, r_0^{2 (2 \ell + 1)} \, \omega^{2 \ell} \, \left(\mathcal{A}^{\ell + 1}_{E_i + \omega} - \mathcal{A}^{\ell + 1}_{E_i - \omega} \right) \, ,
    \end{equation}
    where
    \begin{align}
        \mathcal{A}^{\ell + 1}_{E_i + \omega} - \mathcal{A}^{\ell + 1}_{E_i - \omega} =\, &\frac{\mathcal{B}^{\ell + 1}_{E_i + \omega} \, \sinh\!\left(\!2 \pi \sqrt{\frac{2 (E_i + \omega)}{E_b} - \ell (\ell + 1)} \right) \, \Theta\!\left(\!E_i + \omega - \frac{\ell (\ell + 1)}{2} E_b \right)}{\cosh\!\left(\!2 \pi \sqrt{\frac{2 (E_i + \omega)}{E_b}} \right) - \cosh\!\left(\!2 \pi \sqrt{\frac{2 E_i}{E_b}} \right)}\nonumber\\
        \label{eq:DA_lp1}
        &- \frac{\mathcal{B}^{\ell + 1}_{E_i - \omega} \, \sinh\!\left(\!2 \pi \sqrt{\frac{2 (E_i - \omega)}{E_b} - \ell (\ell + 1)} \right) \, \Theta\!\left(\!E_i - \omega - \frac{\ell (\ell + 1)}{2} E_b \right)}{\cosh\!\left(\!2 \pi \sqrt{\frac{2 E_i}{E_b}} \right) - \cosh\!\left(\!2 \pi \sqrt{\frac{2 (E_i - \omega)}{E_b}} \right)} \, .
    \end{align}
    \item Photon with $\ell = 1$: see equation~\eqref{eq:l1Ph}.
    \item Photon with $\ell \ge 2$:
    \begin{equation}
        \label{eq:sigmaabs_lge2_photon}
        \sigma_\text{abs} = \frac{(\ell - 1) \, (\ell + 1)^2 \, \pi^2 \, \Gamma(\ell - 1)^2}{2^{2 (\ell - 1)} \, \Gamma(2 \ell)^2 \, \Gamma\!\left(\ell + \frac{1}{2} \right)^2} \, E_b^{2 (\ell - 1)} \, r_0^{4 \ell} \, \omega^{2 \ell} \, \left(\mathcal{A}^\ell_{E_i + \omega} - \mathcal{A}^\ell_{E_i - \omega} \right) \, ,
    \end{equation}
    where
    \begin{align}
        \mathcal{A}^{\ell}_{E_i + \omega} - \mathcal{A}^{\ell}_{E_i - \omega} =\, &\frac{\mathcal{B}^\ell_{E_i + \omega} \, \sinh\!\left(\!2 \pi \sqrt{\frac{2 (E_i + \omega)}{E_b} - \ell (\ell + 1)} \right) \, \Theta\!\left(\!E_i + \omega - \frac{\ell (\ell + 1)}{2} E_b \right)}{\cosh\!\left(\!2 \pi \sqrt{\frac{2 (E_i + \omega)}{E_b}} \right) - \cosh\!\left(\!2 \pi \sqrt{\frac{2 E_i}{E_b}} \right)}\nonumber\\
        \label{eq:DA_l}
        &- \frac{\mathcal{B}^\ell_{E_i - \omega} \, \sinh\!\left(\!2 \pi \sqrt{\frac{2 (E_i - \omega)}{E_b} - \ell (\ell + 1)} \right) \, \Theta\!\left(\!E_i - \omega - \frac{\ell (\ell + 1)}{2} E_b \right)}{\cosh\!\left(\!2 \pi \sqrt{\frac{2 E_i}{E_b}} \right) - \cosh\!\left(\!2 \pi \sqrt{\frac{2 (E_i - \omega)}{E_b}} \right)} \, ,
    \end{align}
    The only difference between \eqref{eq:DA_lp1} and \eqref{eq:DA_l} lies in the conformal dimension $\Delta$ appearing in $\mathcal{B}^\Delta_{E_i \pm \omega}$, which differs between the two types of fields.
    \item Graviton with $\ell \ge 2$:
    \begin{equation}
        \label{eq:sigmaabs_graviton}
        \sigma_\text{abs} = \frac{(\ell + 2) \, (\ell + 1)^2 \, \pi^2 \, \Gamma(\ell - 1)^2}{2^{2 (\ell - 1)} \, \Gamma(2 \ell)^2 \, \Gamma\!\left(\ell + \frac{1}{2} \right)^2} \, E_b^{2 (\ell - 1)} \, r_0^{4 \ell} \, \omega^{2 \ell} \, \left(\mathcal{A}^\ell_{E_i + \omega} - \mathcal{A}^\ell_{E_i - \omega} \right) \, ,
    \end{equation}
    where $\mathcal{A}^\ell_{E_i + \omega} - \mathcal{A}^\ell_{E_i - \omega}$ is given explicitly in \eqref{eq:DA_l}.
    \item Dark photon with $\ell \ge 1$:
    \begin{equation}
        \label{eq:sigmaabs_hidden_photon}
        \sigma_\text{abs} = \frac{2^{2 (\ell + 1)} \, \pi \, \Gamma(\ell)^2 \, \Gamma(\ell + 2)^2}{\Gamma(2 \ell + 1)^3 \, \Gamma(2 \ell + 2)} \, E_b^{2 \ell} \, r_0^{2 (2 \ell + 1)} \, \omega^{2 \ell} \, \left(\mathcal{A}^{\ell + 1}_{E_i + \omega} - \mathcal{A}^{\ell + 1}_{E_i - \omega} \right) \, ,
    \end{equation}
    where $\mathcal{A}^{\ell + 1}_{E_i + \omega} - \mathcal{A}^{\ell + 1}_{E_i - \omega}$ is given explicitly in \eqref{eq:DA_lp1}.
\end{itemize}

\section{Quantum vs.~semiclassical transition probabilities and final-state densities}
\label{app:B}

In our setting, where we take $j=0$ for the initial black hole spin, the explicit expressions for the quantum transition probabilities \eqref{eq:f_def} and density of final states \eqref{eq:rho_Ef} are, respectively,
\begin{align}
    \label{eq:f_qu}
    f_\Delta(E_i \pm \omega) &= \pm \, \frac{\pi^2 \, \mathcal{B}^\Delta_{E_i \pm \omega} \, e^{-S_0} \, E_b^{2 \Delta - 1} \, \omega}{2^{2 \Delta - 3} \, \Gamma(2 \Delta) \, \left[\cosh\!\left(2 \pi \, \sqrt{2 \, (E_i \pm \omega)/E_b} \right) - \cosh\!\left(2 \pi \, \sqrt{2 \, E_i/E_b} \right) \right]} \, ,\\
    \label{eq:rho_qu}
    \rho_\ell(E_i \pm \omega) &= \frac{(2 \ell + 1) \, e^{S_0}}{2 \, \pi^2 \, E_b} \, \sinh\!\left(2 \pi \, \sqrt{2 \, (E_i \pm \omega) / E_b - \ell \, (\ell + 1)} \right) \, \Theta\Big(\!2 \, (E_i \pm \omega)/E_b - \ell \, (\ell + 1) \!\Big) \, ,
\end{align}
with $\mathcal{B}^\Delta_{E_i \pm \omega}$ given by \eqref{eq:sBp_1}--\eqref{eq:sBpm_prod}. If we expand them in the semiclassical regime $\omega, E_b \ll E_i$, we find
\begin{align}
    \label{eq:f_sc}
    f_\Delta(E_i \pm \omega) &\approx \pm \frac{2 \, (2 \pi)^{2 \Delta} \, e^{-S_0} \, E_b \, \sinh\!\left(\!\frac{\beta \omega}{2} \!\right) \, \Gamma\!\left(\!\Delta + i \, \frac{\beta \omega}{2 \pi} \!\right) \, \Gamma\!\left(\!\Delta - i \, \frac{\beta \omega}{2 \pi} \!\right)}{\Gamma(2 \Delta) \, \beta^{2 \Delta - 1} \, e^{\frac{(2 \pi)^2}{\beta E_b}} \, \left(e^{\pm \beta \omega} - 1 \right)} \, ,\\
    \label{eq:rho_sc}
    \rho_\ell(E_i \pm \omega) &\approx \frac{(2 \ell + 1) \, e^{S_0}}{(2 \pi)^2 \, E_b} \, e^{\frac{(2 \pi)^2}{\beta E_b} \pm \beta \omega} \, ,
\end{align}
where $\beta$ is the inverse Hawking temperature of the black hole \eqref{eq:T_leading}. In writing the expression for the semiclassical density of states \eqref{eq:rho_sc}, we also assumed $\omega \gg E_b$ since semiclassical physics is insensitive to the quantum energy gap $E_{0,\ell} \sim E_b$. However, note that we do not need to assume $\omega \gg E_b$ in order to obtain the semiclassical greybody factor \eqref{eq:PgbSC}. If we do not make this assumption, then we would have an extra exponential $e^{-\beta E_{0,\ell}}$ in the density of states \eqref{eq:rho_sc}.

%%%%%%%%%%%%%%%%%%%%%%%%%%%%%%%%%%%%%
\bibliography{QCSbib}

\providecommand{\href}[2]{#2}\begingroup\raggedright\begin{thebibliography}{10}

\bibitem{Iliesiu:2020qvm}
L.~V. Iliesiu and G.~J. Turiaci, \emph{{The statistical mechanics of near-extremal black holes}}, \href{https://doi.org/10.1007/JHEP05(2021)145}{\emph{JHEP} {\bfseries 05} (2021) 145} [\href{https://arxiv.org/abs/2003.02860}{{\ttfamily 2003.02860}}].

\bibitem{Mertens:2022irh}
T.~G. Mertens and G.~J. Turiaci, \emph{{Solvable models of quantum black holes: a review on Jackiw\textendash{}Teitelboim gravity}}, \href{https://doi.org/10.1007/s41114-023-00046-1}{\emph{Living Rev. Rel.} {\bfseries 26} (2023) 4} [\href{https://arxiv.org/abs/2210.10846}{{\ttfamily 2210.10846}}].

\bibitem{Bai:2023hpd}
Y.~Bai and M.~Korwar, \emph{{Near-extremal charged black holes: greybody factors and evolution}}, \href{https://doi.org/10.1007/JHEP03(2023)151}{\emph{JHEP} {\bfseries 03} (2023) 151} [\href{https://arxiv.org/abs/2301.07739}{{\ttfamily 2301.07739}}].

\bibitem{Brown:2024ajk}
A.~R. Brown, L.~V. Iliesiu, G.~Penington and M.~Usatyuk, \emph{{The evaporation of charged black holes}},  \href{https://arxiv.org/abs/2411.03447}{{\ttfamily 2411.03447}}.

\bibitem{Maulik:2025hax}
S.~Maulik, X.~Meng and L.~A. Pando~Zayas, \emph{{Quantum-Corrected Hawking Radiation from Near-Extremal Kerr-Newman Black Holes}},  \href{https://arxiv.org/abs/2501.08252}{{\ttfamily 2501.08252}}.

\bibitem{Emparan:2025sao}
R.~Emparan, \emph{{Quantum cross-section of near-extremal black holes}}, \href{https://doi.org/10.1007/JHEP04(2025)122}{\emph{JHEP} {\bfseries 04} (2025) 122} [\href{https://arxiv.org/abs/2501.17470}{{\ttfamily 2501.17470}}].

\bibitem{Biggs:2025nzs}
A.~Biggs, \emph{{Following the state of an evaporating charged black hole into the quantum gravity regime}},  \href{https://arxiv.org/abs/2503.02051}{{\ttfamily 2503.02051}}.

\bibitem{Lin:2025wof}
G.~Lin, L.~V. Iliesiu and M.~Usatyuk, \emph{{The evaporation of black holes in supergravity}},  \href{https://arxiv.org/abs/2504.21077}{{\ttfamily 2504.21077}}.

\bibitem{Li:2025vcm}
R.~Li, Z.-X. Man and J.~Wang, \emph{{Decoherence of quantum superpositions in near-extremal Reissner-Nordstr{\"o}m black holes with quantum gravity corrections}},  \href{https://arxiv.org/abs/2505.07480}{{\ttfamily 2505.07480}}.

\bibitem{Mertens:2019bvy}
T.~G. Mertens, \emph{{Towards Black Hole Evaporation in Jackiw-Teitelboim Gravity}}, \href{https://doi.org/10.1007/JHEP07(2019)097}{\emph{JHEP} {\bfseries 07} (2019) 097} [\href{https://arxiv.org/abs/1903.10485}{{\ttfamily 1903.10485}}].

\bibitem{Blommaert:2020yeo}
A.~Blommaert, T.~G. Mertens and H.~Verschelde, \emph{{Unruh detectors and quantum chaos in JT gravity}}, \href{https://doi.org/10.1007/JHEP03(2021)086}{\emph{JHEP} {\bfseries 03} (2021) 086} [\href{https://arxiv.org/abs/2005.13058}{{\ttfamily 2005.13058}}].

\bibitem{Iliesiu:2022onk}
L.~V. Iliesiu, S.~Murthy and G.~J. Turiaci, \emph{{Revisiting the Logarithmic Corrections to the Black Hole Entropy}},  \href{https://arxiv.org/abs/2209.13608}{{\ttfamily 2209.13608}}.

\bibitem{Kapec:2023ruw}
D.~Kapec, A.~Sheta, A.~Strominger and C.~Toldo, \emph{{Logarithmic Corrections to Kerr Thermodynamics}}, \href{https://doi.org/10.1103/PhysRevLett.133.021601}{\emph{Phys. Rev. Lett.} {\bfseries 133} (2024) 021601} [\href{https://arxiv.org/abs/2310.00848}{{\ttfamily 2310.00848}}].

\bibitem{Rakic:2023vhv}
I.~Rakic, M.~Rangamani and G.~J. Turiaci, \emph{{Thermodynamics of the near-extremal Kerr spacetime}}, \href{https://doi.org/10.1007/JHEP06(2024)011}{\emph{JHEP} {\bfseries 06} (2024) 011} [\href{https://arxiv.org/abs/2310.04532}{{\ttfamily 2310.04532}}].

\bibitem{Emparan:2023ypa}
R.~Emparan and J.~M. Magan, \emph{{Tearing down spacetime with quantum disentanglement}}, \href{https://doi.org/10.1007/JHEP03(2024)078}{\emph{JHEP} {\bfseries 03} (2024) 078} [\href{https://arxiv.org/abs/2312.06803}{{\ttfamily 2312.06803}}].

\bibitem{Maulik:2024dwq}
S.~Maulik, L.~A. Pando~Zayas, A.~Ray and J.~Zhang, \emph{{Universality in logarithmic temperature corrections to near-extremal rotating black hole thermodynamics in various dimensions}}, \href{https://doi.org/10.1007/JHEP06(2024)034}{\emph{JHEP} {\bfseries 06} (2024) 034} [\href{https://arxiv.org/abs/2401.16507}{{\ttfamily 2401.16507}}].

\bibitem{Kapec:2024zdj}
D.~Kapec, Y.~T.~A. Law and C.~Toldo, \emph{{Quasinormal Corrections to Near-Extremal Black Hole Thermodynamics}},  \href{https://arxiv.org/abs/2409.14928}{{\ttfamily 2409.14928}}.

\bibitem{Ferko:2024uxi}
C.~Ferko, S.~Murthy and M.~Rangamani, \emph{{Strings in AdS$_{3}$: one-loop partition function and near-extremal BTZ thermodynamics}}, \href{https://doi.org/10.1007/JHEP05(2025)010}{\emph{JHEP} {\bfseries 05} (2025) 010} [\href{https://arxiv.org/abs/2408.14567}{{\ttfamily 2408.14567}}].

\bibitem{Kolanowski:2024zrq}
M.~Kolanowski, D.~Marolf, I.~Rakic, M.~Rangamani and G.~J. Turiaci, \emph{{Looking at extremal black holes from very far away}},  \href{https://arxiv.org/abs/2409.16248}{{\ttfamily 2409.16248}}.

\bibitem{Arnaudo:2024bbd}
P.~Arnaudo, G.~Bonelli and A.~Tanzini, \emph{{One-loop corrections to near extremal Kerr thermodynamics from semiclassical Virasoro blocks}},  \href{https://arxiv.org/abs/2412.16057}{{\ttfamily 2412.16057}}.

\bibitem{Blacker:2025zca}
M.~J. Blacker, A.~Castro, W.~Sybesma and C.~Toldo, \emph{{Quantum corrections to the path integral of near extremal de Sitter black holes}},  \href{https://arxiv.org/abs/2503.14623}{{\ttfamily 2503.14623}}.

\bibitem{Mariani:2025hee}
F.~Mariani and C.~Toldo, \emph{{Gravitational dynamics of near-extreme Kerr (Anti-)de Sitter black holes}},  \href{https://arxiv.org/abs/2505.02674}{{\ttfamily 2505.02674}}.

\bibitem{Arnaudo:2025btb}
P.~Arnaudo, G.~Bonelli and A.~Tanzini, \emph{{One loop corrections to the thermodynamics of near-extremal Kerr-(A)dS black holes from Heun equation}},  \href{https://arxiv.org/abs/2506.08959}{{\ttfamily 2506.08959}}.

\bibitem{Heydeman:2020hhw}
M.~Heydeman, L.~V. Iliesiu, G.~J. Turiaci and W.~Zhao, \emph{{The statistical mechanics of near-BPS black holes}}, \href{https://doi.org/10.1088/1751-8121/ac3be9}{\emph{J. Phys. A} {\bfseries 55} (2022) 014004} [\href{https://arxiv.org/abs/2011.01953}{{\ttfamily 2011.01953}}].

\bibitem{Danielson:2022tdw}
D.~L. Danielson, G.~Satishchandran and R.~M. Wald, \emph{{Black holes decohere quantum superpositions}}, \href{https://doi.org/10.1142/S0218271822410036}{\emph{Int. J. Mod. Phys. D} {\bfseries 31} (2022) 2241003} [\href{https://arxiv.org/abs/2205.06279}{{\ttfamily 2205.06279}}].

\bibitem{Biggs:2024dgp}
A.~Biggs and J.~Maldacena, \emph{{Comparing the decoherence effects due to black holes versus ordinary matter}},  \href{https://arxiv.org/abs/2405.02227}{{\ttfamily 2405.02227}}.

\bibitem{Crispino:2009zza}
L.~C.~B. Crispino, A.~Higuchi and E.~S. Oliveira, \emph{{Electromagnetic absorption cross section of Reissner-Nordstrom black holes revisited}}, \href{https://doi.org/10.1103/PhysRevD.80.104026}{\emph{Phys. Rev. D} {\bfseries 80} (2009) 104026}.

\bibitem{Oliveira:2011zz}
E.~S. Oliveira, L.~C.~B. Crispino and A.~Higuchi, \emph{{Equality between gravitational and electromagnetic absorption cross sections of extreme Reissner-Nordstrom black holes}}, \href{https://doi.org/10.1103/PhysRevD.84.084048}{\emph{Phys. Rev. D} {\bfseries 84} (2011) 084048}.

\bibitem{Crispino:2000jx}
L.~C.~B. Crispino, A.~Higuchi and G.~E.~A. Matsas, \emph{{Quantization of the electromagnetic field outside static black holes and its application to low-energy phenomena}}, \href{https://doi.org/10.1103/PhysRevD.80.029906}{\emph{Phys. Rev. D} {\bfseries 63} (2001) 124008} [\href{https://arxiv.org/abs/gr-qc/0011070}{{\ttfamily gr-qc/0011070}}].

\bibitem{Page:2000dk}
D.~N. Page, \emph{{Thermodynamics of near extreme black holes}},  \href{https://arxiv.org/abs/hep-th/0012020}{{\ttfamily hep-th/0012020}}.

\bibitem{Arbey:2021jif}
A.~Arbey, J.~Auffinger, M.~Geiller, E.~R. Livine and F.~Sartini, \emph{{Hawking radiation by spherically-symmetric static black holes for all spins: Teukolsky equations and potentials}}, \href{https://doi.org/10.1103/PhysRevD.103.104010}{\emph{Phys. Rev. D} {\bfseries 103} (2021) 104010} [\href{https://arxiv.org/abs/2101.02951}{{\ttfamily 2101.02951}}].

\bibitem{Arbey:2021yke}
A.~Arbey, J.~Auffinger, M.~Geiller, E.~R. Livine and F.~Sartini, \emph{{Hawking radiation by spherically-symmetric static black holes for all spins. II. Numerical emission rates, analytical limits, and new constraints}}, \href{https://doi.org/10.1103/PhysRevD.104.084016}{\emph{Phys. Rev. D} {\bfseries 104} (2021) 084016} [\href{https://arxiv.org/abs/2107.03293}{{\ttfamily 2107.03293}}].

\end{thebibliography}\endgroup

\end{document}